\documentclass{article}

\usepackage{arxiv}

\usepackage[utf8]{inputenc} 
\usepackage[T1]{fontenc}    

\usepackage{booktabs}       
\usepackage{amsfonts}       
\usepackage{amsmath}
\usepackage{nicefrac}       
\usepackage{microtype}      
\usepackage{lipsum}		
\usepackage{graphicx}
\usepackage[numbers,compress]{natbib}
\usepackage{doi}
\usepackage{enumitem}
\usepackage{subcaption}

\def\gder#1#2{\frac{\bar{\partial}#1}{\partial #2}}
\def\gtder#1#2{\frac{\bar{\mathrm{d}}#1}{\mathrm{d} #2}}
\def\gmder#1#2#3{\frac{\bar{\partial}^{#1}#2}{\partial #3^{#1}}}
\def\gmtder#1#2#3{\frac{{\bar{\mathrm{d}}^{#1}#2}}{\mathrm{d} #3^{#1}}}

\def\pmder#1#2#3{\frac{\partial^{#1}#2}{\partial #3}}
\def\vect#1{\mathbf{#1}}
\def\vecg#1{\boldsymbol{#1}}

\def\SI#1#2{\ensuremath{#1\hspace{1.5pt}\text{#2}}}
\def\unexp#1#2{\ensuremath{\text{#1}^{#2}}}

\def\Hz{Hz}
\def\kg{kg}

\def\m{m}
\def\N{N}

\def\s{s}

\title{Free and forced vibrations of damped locally-resonant sandwich beams}

\author{Andrea Francesco Russillo\\
	Department of Civil, Environmental,\\ 
	Energy and Materials Engineering (DICEAM)\\ 
	University of Reggio Calabria\\ 
	Via Graziella, 89124 Reggio Calabria, Italy\\ \texttt{andreaf.russillo@unirc.it} \\
	\And
	Giuseppe Failla\\
	Department of Civil, Environmental,\\ 
	Energy and Materials Engineering (DICEAM)\\ 
	University of Reggio Calabria\\ 
	Via Graziella, 89124 Reggio Calabria, Italy\\ \texttt{giuseppe.failla@unirc.it} \\
	\AND
	Fernando Fraternali\\
	Department of Civil Engineering\\
	 University of Salerno\\
	 84084 Fisciano (Salerno), Italy\\ \texttt{f.fraternali@unisa.it}}

\date{}

\renewcommand{\headeright}{}
\renewcommand{\undertitle}{}
\renewcommand{\shorttitle}{Free and forced vibrations of damped locally-resonant sandwich beams}

\begin{document}
\maketitle

\begin{abstract}
This paper addresses the dynamics of locally-resonant sandwich beams, where multi-degree-of-freedom viscously-damped resonators are periodically distributed within the core matrix. Using an equivalent single-layer Timoshenko beam model coupled with mass-spring-dashpot subsystems representing the resonators, two solution methods are presented. The first is a direct integration method providing the exact frequency response under arbitrary loads. The second is a complex modal analysis approach obtaining exact modal impulse and frequency response functions, upon deriving appropriate orthogonality conditions for the complex modes. The challenging issue of calculating all eigenvalues, without missing anyone, is solved applying a recently-introduced contour-integral algorithm to a characteristic equation built as determinant of an exact frequency-response matrix, whose size is $4 \times 4$ regardless of the number of resonators. Numerical applications prove exactness and robustness of the proposed solutions. 
\end{abstract}


\section{Introduction}
The concept of locally-resonant beam is an emerging concept in engineering. It defines a beam with periodically-attached resonators, where periodicity and local resonance ensure inherent attenuation properties of elastic waves over frequency bands named \textit{band gaps}. Depending on the dynamic properties (mass/stiffness) and mutual distance of the resonators, the band gaps may fall well below the Bragg frequency, providing remarkable vibration mitigation effects in several engineering problems. On the other hand, experimental evidence confirmed that the dynamics of locally-resonant beams can be accurately predicted by relatively-simple computational models, involving Euler-Bernoulli or Timoshenko continuous beams coupled with mass-spring subsystems representing the resonators. Several studies supported by experimental results have been published in this respect \cite{xiao2012broadband,xiao2013flexural,xiao2013theoretical,sun2010theory,zhu2014chiral,
pai2010metamaterial,hu2018internally,casalotti2018metamaterial,wang2016multi,
chen2011containing,hussein2014dynamics,sugino2017general,baravelli2013internally,
beli2018wave,zhou2019actively,liu2020study,krushynska2017coupling,miniaci2016spider}.

Recently, the concept of locally-resonant beam has been proposed also for sandwich beams, which are ideally suitable to host small resonators within the core matrix, featuring single or multiple degrees of freedom (DOFs). Pioneering work in this field is due to Sun and co-workers \cite{chen2011containing,chen2011dynamic,sharma2016local,sharma2016impact,chen2012reducing,
chen2016sandwich,chen2013wave}. They proposed an equivalent single-layer Timoshenko beam model coupled with mass-spring subsystems representing the resonators, investigating the dynamic behaviour under different excitations, including impact \cite{sharma2016impact} and moving ones \cite{chen2016sandwich}. The mass-spring subsystems were considered as exerting point forces \cite{chen2011dynamic,sharma2016local} or distributed forces over the mutual distance \cite{chen2011containing,chen2011dynamic}. The equivalent single-layer Timoshenko beam model was validated numerically by comparison with finite-element models detailing the various layers of the beam \cite{chen2011dynamic}, while the wave attenuation properties were confirmed by experimental tests \cite{chen2011containing}. Other authors are currently working on alternative concepts of locally-resonant sandwich beams, including periodic viscoelastic core matrices \cite{guo2017flexural}, lattice truss cores \cite{li2020phononic} and meta-lattice resonant truss cores \cite{li2019novel}.     

This paper aims to contribute to the study of locally-resonant sandwich beams hosting small multi-DOF re\-so\-na\-tors within the core matrix \cite{chen2011containing,chen2011dynamic,sharma2016local,sharma2016impact,
chen2012reducing,chen2016sandwich,chen2013wave}, focusing on free and forced vibrations in presence of viscous damping within the resonators. Modelling the system as an equivalent single-layer Timoshenko beam coupled with mass-spring-dashpot subsystems exerting transverse point forces \cite{chen2011dynamic,sharma2016local}, two solution methods are introduced. First, a direct integration method provides the exact frequency-response in analytical form for arbitrary loads, based on a direct/inverse Laplace Transform of the motion equations. Second, the modal impulse and frequency responses are obtained by a complex modal analysis approach, being damping not proportional. In this context, the main difficulty is the calculation of all complex eigenvalues without missing anyone as indeed, in presence of viscous damping, the well-established Wittrick-Williams algorithm \cite{williams1970automatic,qi2004accurate} is no longer applicable to calculate all roots of the characteristic equation. This challenge is successfully solved by a suitable contour-integral algorithm recently introduced for general nonlinear eigenvalue problems \cite{sakurai2003projection,asakura2009numerical,ikegami2010filter} and applied, in this paper, to a characteristic equation built as determinant of an exact frequency response matrix of size $4 \times 4$, regardless of the number of resonators;  to the best of authors' knowledge, this is the first application of the algorithm \cite{sakurai2003projection,asakura2009numerical,ikegami2010filter} in this context. Finally, once the eigenvalues are calculated, the sought exact modal impulse and frequency responses are built in analytical form, upon deriving orthogonality conditions pertinent to the complex modes.
 
The paper is organized as follows. On introducing the fundamental equations of the locally-resonant sandwich beams under study in Section 2, the direct integration method providing the exact frequency response is described in Section 3, while the complex modal analysis approach is discussed in Section 4. Numerical applications are reported in Section 5. Two Appendices are included. Appendix~A reports details on equations in Section 3; Appendix~B shows how the proposed solution methods can readily be generalized to consider the mass-spring-dashpot subsystems as exerting distributed forces, as in ref.~\cite{chen2011containing,chen2011dynamic}.

\section{Problem under study}
\begin{figure*}
\centering
\includegraphics[scale=0.95]{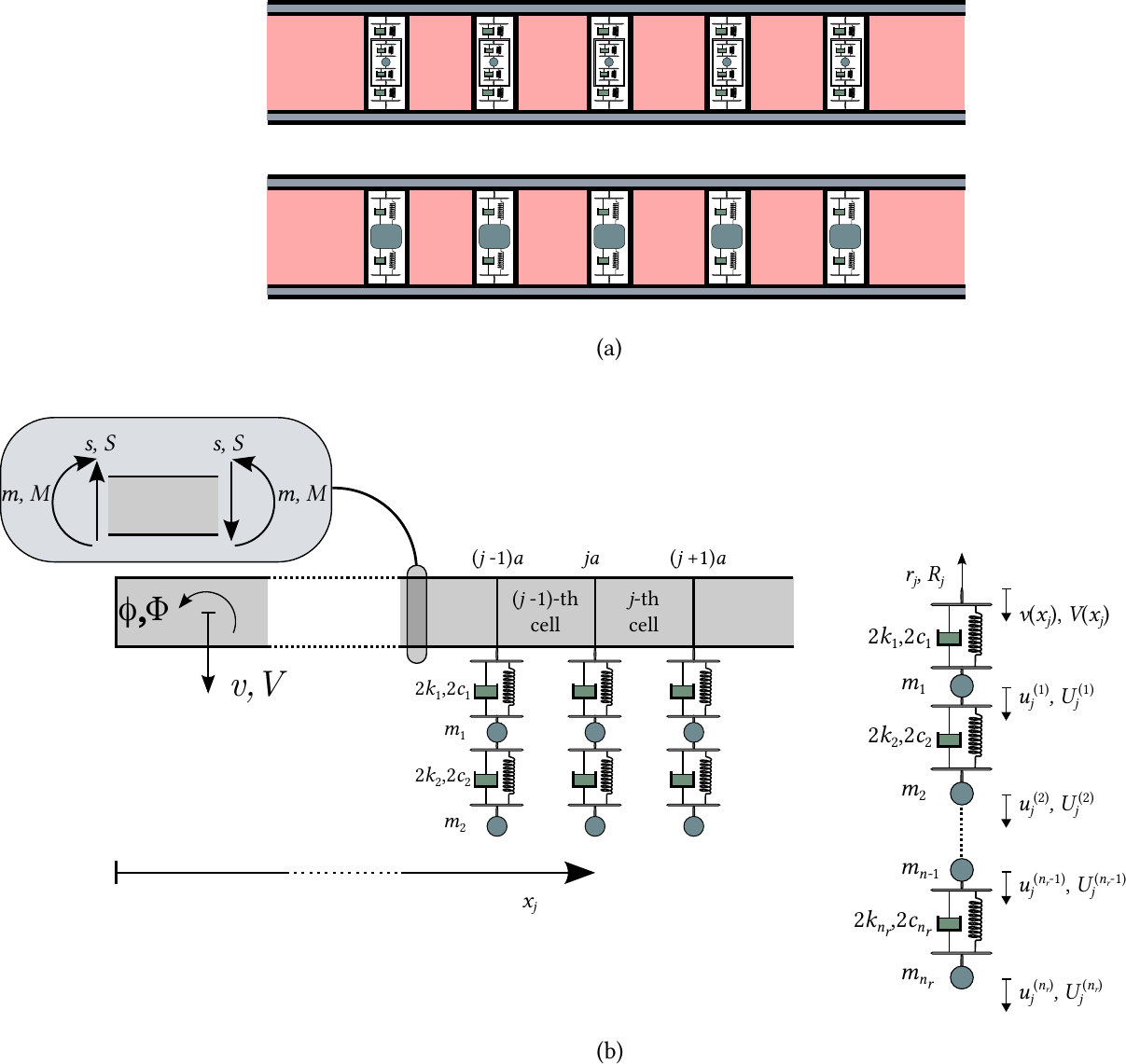}
\caption{\label{fig:figure1}Locally-resonant sandwich beam: (a) 1-DOF or 2-DOF resonators; (b) equivalent single-layer Timoshenko beam model coupled with mass-spring-dashpot resonators ($n_r$ degrees of freedom, for generality)}
\end{figure*}
Consider the locally-resonant sandwich beam in Figure \ref{fig:figure1}, consisting of two thin face-sheets applied below and above a thick core material hosting periodically-distributed re\-so\-na\-tors. Every resonator may include one or multiple masses, connected to each other and to the beam by linearly-elastic springs and viscous dashpots, as shown in Figure~\ref{fig:figure1}a. 
Following ref. \cite{chen2011dynamic,sharma2016local}, the system is represented as an equivalent single-layer Timoshenko beam coupled with mass-spring-dashpot subsystems exerting point forces, as shown in Figure~\ref{fig:figure1}b. The two equations of motion under a dynamic transverse load read:
\begin{small}
\begin{flalign}
&GA\left(\gmder{2}{v}{x}+\gder{\phi}{x}\right)-\rho A \pmder{2}{v}{t} + \sum_{j=1}^{N}r_j \delta(x-x_j) + p_{v} = 0\label{eqn:const_1}&&\\ 
&EI\gmder{2}{\phi}{x}- GA\left(\gder{v}{x}+\phi\right)-\rho I\pmder{2}{\phi}{t} + p_{\phi} = 0&&\label{eqn:const_2} 
\end{flalign}
\end{small}

\noindent
where $v=v(x,t)$ and $\phi=\phi(x,t)$ are deflection (positive downward) and rotation (positive counterclockwise) of the cross section, $p_{v}=p_{v}(x,t)$ and $p_{\phi} = p_{\phi}(x,t)$ are the transversal and rotational dynamical load respectively; bar means generalized derivative and symbol $\delta(x-x_j)$ denotes a Dirac's delta at $x_j$; further, symbols $r_j = r_j(t)$  and $x_j =ja$  are reaction force and application point of the $j^{\mathrm{th}}$ resonator for $j=1,...,N$, while symbol $a$ denotes the mutual distance. The beam parameters in Eqs.~\eqref{eqn:const_1}-\eqref{eqn:const_2} are calculated in ref. \cite{chen2011dynamic,sharma2016local} as:
\begin{align}
&EI = E_{f}b(h^{2}_c h_f/2 +h_c h_{f}^2)\\
&GA = G_c b (h_c + 2h_f)\\
&\rho A = 2\rho_f b h_f + \rho_c b h_c \\
&\rho I = \rho_f b (h_c^2  h_f/2 + h_c h_f^2)+ \rho_c b h_c^3 /12
\end{align}
where $b = \text{width}$, $h = \text{thickness}$, $E = \text{elastic modulus}$, $G = \text{shear modulus}$, $\rho = \text{mass density}$, while subscripts ``$c$'' and ``$f$'' stand for \textit{core} and \textit{face-sheet} respectively. Validation for the equivalent single-layer Timoshenko beam model governed by Eqs.~\eqref{eqn:const_1}-\eqref{eqn:const_2} has been provided in ref.~\cite{chen2011dynamic} by comparison with finite-element models detailing the various layers of the beam. 

The vibration response can be represented as
\begin{equation}\label{eqn:vibration_response}
\vect{y}=\vect{Y}\mathrm{e}^{\mathrm{i}\omega t}; \quad \vect{u}_j=\vect{U}_j\mathrm{e}^{\mathrm{i}\omega t}
\end{equation}
where $\vect{y} = \vect{y}(x,t) = \begin{bmatrix}
v&\phi&m&s
\end{bmatrix}^{\text{T}}$ and~~$\vect{Y} = \Big[
V~~~\Phi~~~M\\S
\Big]^{\text{T}}$, $\vect{u}_j = \vect{u}_j (t) =  \begin{bmatrix}
u_j^{(1)}&u_j^{(2)}&\dots&u_j^{(n_{r})}
\end{bmatrix}^{\text{T}}$ and~~$\vect{U}_j =  \begin{bmatrix}
U_j^{(1)}&U_j^{(2)}&\dots&U_j^{(n_{r})}
\end{bmatrix}^{\text{T}}$  collect the response variables of the beam and the resonator applied at $x = x_j$. Eq.~\eqref{eqn:vibration_response} is a general form to represent:
\begin{enumerate}[label=\alph*)]
\item Frequency response under an harmonic load with frequency $\omega$, i.e. $\vect{Y}=\vect{Y}(x,\omega)$ and $\vect{U}_j=\vect{U}_j(\omega)$;
\item Free-vibration response, being $\omega = \omega_k$ an eigenvalue and $\vect{Y} =\vect{Y}_k (x)$, $\vect{U}_{j} = \vect{U}_{j,k}$ the corresponding eigenfunctions. Eigenvalues and eigenfunctions are generally complex as viscous dashpots within the resonators make damping not proportional.
\end{enumerate}
Alternative equations for the sandwich beam in Figure~\ref{fig:figure1} were provided in ref.~\cite{chen2011containing,chen2011dynamic}, where the reaction force of every resonator is taken as a distributed force over the mutual distance $a$. This model is not treated in details here; however, it can be handled with little changes to the solutions proposed in Sections 3-4, as explained in Appendix~B of the paper.

\section{Frequency response}\label{sec:fr_response}
Be the beam in Figure~\ref{fig:figure1} subjected to a harmonic load $p_{v}(x,t)=f_{v}(x)\mathrm{e}^{\mathrm{i}\omega t}$ and $p_{\phi}(x,t)=f_{\phi}(x)\mathrm{e}^{\mathrm{i}\omega t}$, so that the frequency response takes the form~\eqref{eqn:vibration_response}. Using the theory of generalized functions \cite{falsone2002use,caddemi2013exact,biondi2007euler,burlon2016exact,di2018flexural}, the equations of motion in the frequency domain read:
\begin{small}
\begin{flalign}
&GA\left(\gmtder{2}{V}{x}+\gtder{\Phi}{x}\right)+\rho A \omega^2 V + \sum_{j=1}^{N}R_j \delta(x-x_j) + f_{v} = 0&& \label{eqn:disp_freq}\\
&EI\gmtder{2}{\Phi}{x}- GA\left(\gtder{V}{x}+\Phi\right)+\rho I \omega^2 \Phi + f_{\phi}= 0 && \label{eqn:rot_freq}
\end{flalign}
\end{small}

\noindent
where $R_j$ is the reaction force of the $j^{\mathrm{th}}$ resonator, given as
\begin{equation}\label{eqn:resonator_force}
R_j = -k_{eq}(\omega)V(x_j)
\end{equation}
In Eq.~\eqref{eqn:resonator_force} $k_{eq}(\omega)$ is the frequency-dependent stiffness of the resonator, which can be obtained from its equations of motion in the frequency domain. Specifically, for a chain of masses, springs and dashpots as in Figure~\ref{fig:figure1}, $k_{eq}(\omega)=k_{eq,j}(\omega)$:
\begin{equation}\label{eqn:equivalent_DS}
k_{eq}(\omega) = \vect{d}_{vu}^{\mathrm{T}}\vect{D}^{-1}_{uu}\vect{d}_{uv} - D_{11}
\end{equation}
having partitioned the dynamic stiffness matrix $\vect{D}_{r}(\omega)$ of the resonator as:
\begin{equation}\label{eqn:dsm_partion}
\vect{D}_{r}(\omega) = \begin{bmatrix}
D_{vv}&\vect{d}_{vu}^{\mathrm{T}}\\
\vect{d}_{uv}&\vect{D}_{uu}
\end{bmatrix}
\end{equation}
where the subscripts $v$ and $u$ are associated, respectively, with the deflection of the resonator application point and the DOFs within the resonator.

Next, the solution is constructed observing that Eqs.~\eqref{eqn:disp_freq}-\eqref{eqn:rot_freq} can be reduced to two decoupled $4^{\mathrm{th}}$ order differential equation only, in the following form:
\begin{equation}\label{eqn:single_motion}
\gmtder{4}{Z}{x} + p_1\gmtder{2}{Z}{x} + p_2 Z + p_3 = 0
\end{equation}
where symbol $Z$ may denote either the deflection $V$ or the rotation $\Phi$, $p_1$, $p_2$ and $p_3$ are given as
\begin{align}
&p_1 =  \left(\frac{\rho EI \omega^2}{G} + \rho I\omega^2 \right)/EI \label{eqn:p1}\\
&p_2 = \left(\frac{\rho^2  I \omega^4}{G}  - \rho A \omega^2\right)/EI \label{eqn:p2}
\end{align}
\begin{equation}
p_3 = 
\left\{
\begin{aligned}
&-\frac{EI}{GA}\gmtder{2}{q}{x} - \left( \frac{\rho I \omega^2}{GA} - 1\right)q +\gtder{f_\phi}{x}& \quad&\text{if}\, Z=V\\
& -\left(\gmtder{2}{f_{\phi}}{x} + \frac{\rho \omega^2}{G}f_{\phi} + \gtder{q}{x}\right)& \quad&\text{if}\, Z=\Phi
\end{aligned}
\right.
\end{equation}
being 
\begin{equation}\label{eqn:load_q}
q = \sum_{j=1}^{N}R_{j}\delta(x-x_j) + f_{v}
\end{equation}
The solution of Eq.~\eqref{eqn:single_motion} takes the expression
\begin{equation}\label{eqn:solution_motion}
Z = Z_{om} + \sum_{j=1}^{N}R_jJ_{Z,j} + X_{Z}^{(f)}
\end{equation}
In Eq.~\eqref{eqn:solution_motion}, $Z_{om}$ is the solution of the homogeneous differential equation associated with Eq.~\eqref{eqn:single_motion}, i.e.
\begin{equation}\label{eqn:homogeneous_solution}
Z_{om} = \sum_{i=1}^{4}c_i \alpha_i \mathrm{e}^{\lambda_i x}
\end{equation}
where $c_i$ are integration constants, $\lambda_i$ are the roots of the characteristic polynomial
\begin{align}\label{eqn:polynomial_roots}
&p_{1,2} = \mp \left(\left(-p_1 - \sqrt{p_1^2 - 4p_2}\right)/2\right)^{1/2}\\
&p_{3,4} = \mp \left(\left(-p_1 + \sqrt{p_1^2 - 4p_2}\right)/2\right)^{1/2}
\end{align}
and the coefficients $\alpha_i$ are
\begin{equation}\label{eqn:alpha}
\alpha_i = \left\{
\begin{aligned}
&1& \quad&\text{if}\, Z = \Phi\\
&-\frac{G A \lambda_i}{\rho A \omega^2 + GA \lambda_i^2}&\quad&\text{if}\, Z = V
\end{aligned}
\right.
\end{equation}
Further in Eq.~\eqref{eqn:solution_motion}, $J_{Z,j}= J_{Z}(x,x_j)$ is the particular integral associated with the Dirac's delta $\delta(x-x_j)$ in Eq.~\eqref{eqn:load_q}, obtained by applying direct and Laplace Transform to Eq.\\~\eqref{eqn:single_motion}, as in ref.~\cite{wang2007vibration} for $Z=V$ (deflection) and $Z=\Phi$ (rotation). Specifically $J_{V}(x,x_j)$ is:
\begin{equation}\label{eqn:fundamental}
J_{Z}(x,x_j) = -(\sqrt{2} GA \Sigma_1)^{-1}[B\sinh (C(x-x_j))+ D\sinh (E(x-x_j))]\mathcal{H} (x-x_j)
\end{equation}
\begin{equation}
\begin{aligned}
&B = \left(\sqrt{2}C\right)^{-1}\left[\Sigma_1 + \Sigma_2 -2 (GA)^2\right]\\
&C = \left(\left(\Sigma_1 - \Sigma_3\right)/(2 EI~GA)\right)^{1/2}\\
&D = -\left(\sqrt{2}C\right)^{-1}\left[ \Sigma_1 - \Sigma_2 +2 (GA)^2\right]\\
&E = \left(-\left(\Sigma_1 + \Sigma_3\right)/(2 EI~GA)\right)^{1/2}\\ 
&\Sigma_1 = \left[(EI)^2 \rho x_0^2 \omega ^4+2 EI GA \rho x_0 \omega ^2 \left(2 GA-I\rho  \omega ^2\right) \right.\\
&\left.+(GA)^2 I^2 \rho ^2 \omega ^4\right]^{1/2}\\
&\Sigma_2 = GA I \rho  \omega ^2 -EI \rho x_0 \omega ^2 \quad \Sigma_3 = GA I \rho  \omega ^2 +EI \rho x_0 \omega ^2 
\end{aligned}
\end{equation}
being $\mathcal{H}(x)$ the unit-step function defined as
\begin{equation}\label{eqn:heaviside}
\mathcal{H}(x) = \left\{
\begin{aligned}
&1&\quad&\text{if}~x>0\\
&0&\quad&\text{if}~x<0\\
\end{aligned}
\right.
\end{equation}
Also, $J_{\Phi}(x,x_j)$ is:
\begin{equation}\label{eqn:fundamental_rotation}
J_{\Phi}(x,x_j) = -GA\Upsilon_1^{-1}\{\cosh [S_1(x-x_j)]-\cosh [S_2 (x-x_j)]\}\mathcal{H}(x-x_j) 
\end{equation}
\begin{equation}\label{eqn:coeff_rotation}
\begin{aligned}
&S_1 = \left((\Upsilon_1 -\Upsilon_2)/(2 EI~GA)\right)^{1/2}\\
&S_2 = \left(-(\Upsilon_1 +\Upsilon_2)/(2 EI~GA)\right)^{1/2}\\
&\Upsilon_1 = \left\{\rho  \omega ^2 \left[(A~EI)^2 \rho  \omega^2+ 2 A EI GA \left(2 GA-I \rho  \omega ^2\right)\right]\right.\\
&\left.+(GA~I)^2 \rho  \omega ^2\right\}^{1/2}\\
&\Upsilon_2 =A EI \rho  \omega ^2+GA I \rho  \omega ^2
\end{aligned}
\end{equation}
Finally, $X_{Z}^{(f)}$ is the particular integral associated with the loads $f_v$ and $f_\phi$, which can be expressed using Eq.~\eqref{eqn:fundamental}
\begin{equation}\label{eqn:particular_load}
X_{Z}^{(f)} = \int_{0}^{L}J_{Z}(x,\xi)f_v(\xi)\,\text{d}\xi + \int_{0}^{L}J_{Z}(x,\xi)f_\phi(\xi)\,\text{d}\xi
\end{equation}

Now, using $Z=V$ given by Eq.~\eqref{eqn:solution_motion} for $V(x_j)$ in Eq.~\eqref{eqn:resonator_force} it is seen that every reaction force $R_j$ depends only on the four integration constants $c_j$ and the reaction forces $R_k$ at $x_k < x_j$. That is, all the reaction forces $R_j$ can be expressed in terms of the integration constants $c_i$, to finally obtain the following expression for the frequency response function (FRF) vector $\vect{Y}(x,\omega)$
\begin{equation}\label{eqn:exact_FRF}
\vect{Y}(x,\omega) = \vect{W}(x,\omega)\vect{c} + \vect{Y}^{(f)}(x,\omega)
\end{equation}
In Eq.~\eqref{eqn:exact_FRF}, $\vect{W}$ is a $4 \times 4$ matrix depending on the solution of the homogeneous equation associated with Eq.~\eqref{eqn:single_motion}, while $\vect{Y}^{(f)}$ is a $4 \times 1$ load-dependent vector. Elements in $\vect{W}$ and $\vect{Y}^{(f)}$ are available in an exact analytical form, and details are given in Appendix~A for conciseness.
Vector $\vect{c}$ in Eq.~\eqref{eqn:exact_FRF} is obtained enforcing the beam boundary conditions (B.C.), i.e.
\begin{equation}\label{eqn:BC_enforcing}
\vect{B}\vect{c} = \vect{e} \quad \rightarrow \quad \vect{c} = \vect{B}^{-1}\vect{e}
\end{equation}
where $\vect{B}$ and $\vect{e}$ are a $4 \times 4$ matrix and $4 \times 1$ vector, built from $\vect{W}$ and $\vect{Y}^{(f)}$ computed at $x = 0$ and $x = L$. The inverse matrix $\vect{B}^{-1}$ is available in a closed analytical form, as shown in ref.~\cite{failla2016exact}. Hence, replacing Eq.~\eqref{eqn:BC_enforcing} for $\vect{c}$ in Eq.~\eqref{eqn:exact_FRF} provides a closed analytical expression for the frequency response vector $\vect{Y}(x,\omega)$ of the beam in Figure~\ref{fig:figure1}, readily implementable in any software package. 

{\sloppy
Now, a few remarks are in order. First, Eq.~\eqref{eqn:exact_FRF} for $\vect{Y}(x,\omega)$ holds for any number of resonators along the beam; resonators applied at the beam ends can be considered as internal resonators located at $x=0^{+}$ and/or $x=L^{-}$ and the corresponding B.C. can be treated as homogeneous. Second, it is noteworthy that the frequency response $\vect{U}_{j}(\omega)$ in the $j^{\mathrm{th}}$ resonator can be obtained from the deflection $V(x_j)$ of the application point, e.g. using the resonator equations of motion in the frequency domain.

A further remark is that Eq.~\eqref{eqn:exact_FRF} can be applied to calculate the transmittance of a cantilever beam \cite{hu2018internally,liu2007design}. In this case, being  $V_{g}\mathrm{e}^{\mathrm{i}\omega t}$ the harmonic deflection at the clamped end, e.g. at $x=0$, $V(x)$ in Eqs.\eqref{eqn:disp_freq}-\eqref{eqn:rot_freq} is the beam deflection relative to the ground and a uniformly-distributed transverse load $f_v = \rho A \omega^2 V_g$ is considered in Eq.~\eqref{eqn:disp_freq}; accordingly, the reaction force  of every resonator shall be set equal to
\begin{equation}\label{eqn:reaction_trasmittance}
R_{j} = -k_{eq}(\omega)(V(x_j) + V_{g})
\end{equation}
while the B.C. are
\begin{equation}\label{eqn:BC_trasmittance}
V(0) = 0 \quad \Phi(0) = 0 \quad M(L) = 0 \quad T(L) = 0
\end{equation}
Based on Eq.~\eqref{eqn:reaction_trasmittance}, changes to matrix $\vect{W}$ and vector $\vect{Y}^{(f)}$ reported in Appendix~A for Eq.~\eqref{eqn:exact_FRF} are straightforward. The transmittance is given as $|V(L) + V_g|/V_g$.
}
Finally, it is noteworthy that the proposed solution \eqref{eqn:exact_FRF} can be readily generalized with little modifications to consider mass-spring-dashpot subsystems as exerting distributed forces over the mutual distance \cite{chen2011containing,chen2011dynamic}; details are given in Appendix~B for brevity.

\section{Modal analysis}\label{sec:modal_analysis}
Damping of the locally-resonant sandwich beam in Figure~\ref{fig:figure1} is generally not proportional. Therefore, a complex modal analysis is required to calculate the eigenvalues with the associated eigenfunctions. Here, the interest is twofold: (1) to use a robust and efficient algorithm to calculate all eigenvalues without missing anyone; (2) to introduce orthogonality conditions pertinent to the single-layer Timoshenko beam model coupled with mass-spring-dashpot resonators, in order to obtain analytical expressions for the modal impulse and frequency response functions. Details will be given next.

\subsection{Calculation of eigenvalues/eigenvectors}\label{sec:countour_integral}

The complex eigenvalues are calculated as the roots of the following characteristic equation obtained from Eq.~\eqref{eqn:BC_enforcing} in free vibrations, i.e. for  $\vect{e}=\vect{0}$:
\begin{equation}\label{eqn:frequencies_dsm}
\det(\vect{B}(\omega)) = 0
\end{equation} 
Eq.~\eqref{eqn:frequencies_dsm} is a transcendental equation and finding all its roots poses computational difficulties, as is typical the case when damped structures are treated by exact dynamic sub-struc\-turing. There exist some methods in the literature to solve characteristic equations derived from a transfer matrix or a dynamic stiffness matrix approach: transfer-matrix based algorithms reverting the zero search to a minimization problem were developed and applied to rods coupled with discrete masses \cite{bestle2014recursive}; further, for 2D frames with viscous beam-column connections, approximate roots were built expanding the frame global dynamic stiffness matrix in series with respect to the circular frequency $\omega$, and neglecting terms higher than the third one \cite{kawashima1984vibration}. 

For the locally-resonant sandwich beam under study, calculating the roots of Eq.~\eqref{eqn:frequencies_dsm} with the required accuracy and without missing anyone is a particularly challenging task because, as a result of local resonance, several modes are expected to exhibit eigenvalues close to each other. Here, the issue is solved using a contour-integral algorithm, recently introduced in the literature for nonlinear eigenvalue problems \cite{sakurai2003projection,asakura2009numerical,ikegami2010filter}.

The contour-integral algorithm requires the dynamic stiffness matrix of the system $\vect{D}(\omega)$ that, for the beams under study, can be readily built using Eq.~\eqref{eqn:exact_FRF}, e.g. following the procedure in ref.~\cite{wang2007vibration}. Specifically, the size of $\vect{D}(\omega)$ is $4\times 4$ for any number $N$ of resonators and any number of DOFs within every resonator. Then, the fundamental steps to calculate the eigenvalues are \cite{sakurai2003projection,asakura2009numerical,ikegami2010filter}:
\begin{enumerate}
\item Selection of a circle $\Gamma = \gamma_{0} + \rho_{0} \mathrm{e}^{\mathrm{i}\theta}$ on the complex plane with center $\gamma_{0}$, radius $\rho_{0}$ and $0\leq \theta\leq 2\pi$.
\item Computation of two complex random source matrices $\vect{U}$ and $\vect{V}$ with dimensions $n_{0} \times L_{0}$, where $n_{0}$ is the size of the dynamic stiffness matrix $\vect{D}(\omega)$ and $L_0$ is the number of source vectors collected in $\vect{U}$ and $\vect{V}$.
\item Computation of the shifted and scaled moments $\vect{M}_{k}$ using $N_{0}$-point trapezoidal rule:
\[
\begin{aligned}
&\vect{S}_k = \frac{1}{N_{0}}\sum_{j=0}^{N_{0}-1}\left(\frac{\omega_j -\gamma}{\rho_0}\right)^{k+1} \vect{D}(\omega_j)^{-1}\vect{V}, \quad k=0,1,...,2K-1\\
&\vect{M}_k = \vect{U}^{\mathrm{H}}\vect{S}_k
\end{aligned}
\]
where $K$ is the maximum moment degree considered for the moment and $\vect{U}^{\mathrm{H}}$ is the Hermitian transpose of $\vect{U}$.
\item Construction of the Hankel matrices $\hat{\vect{H}}_{KL_{0}}$ and \\$\hat{\vect{H}}_{KL_{0}}^{<} \in \mathbb{C}^{KL_{0} \times KL_{0}}$ such that:
\begin{flalign*}
\hat{\vect{H}}_{KL_{0}} = [\vect{M}_{i+j-2}]_{i,j=1}^{K} \quad \hat{\vect{H}}_{KL_{0}} = [\vect{M}_{i+j-1}]_{i,j=1}^{K} &&
\end{flalign*}
\item Perform a singular value decomposition of $\hat{\vect{H}}_{KL_{0}}$.
\item Omit small singular value components $\sigma_i < \epsilon \cdot \max_{i}\sigma_i$, set $\tilde{m}$ as the number of remaining singular value components $(\tilde{m} < KL_{0})$ and construct $\hat{\vect{H}}_{\tilde{m}}$ and $\hat{\vect{H}}_{\tilde{m}}^{<}$ extracting the principal submatrix with maximum index $\tilde{m}$ from $\hat{\vect{H}}_{KL_{0}}$ and $\hat{\vect{H}}_{KL_{0}}^{<}$, that is
\begin{small}
\begin{flalign*}
\hat{\vect{H}}_{\tilde{m}} = \hat{\vect{H}}_{KL_{0}}(1:\tilde{m},1:\tilde{m}); \quad  \hat{\vect{H}}^{<}_{\tilde{m}} = \hat{\vect{H}}^{<}_{KL_{0}}(1:\tilde{m},1:\tilde{m})&&
\end{flalign*}
\end{small}
\item Compute the eigenvalues $\zeta_j$ of the linear pencil:
\[
\hat{\vect{H}}_{\tilde{m}}^{<} = \zeta \hat{\vect{H}}_{\tilde{m}}
\]
\item Calculate the eigenvalues
\[
\omega_j = \gamma_0 + \rho_0 \zeta_j, \quad j = 1,...,\tilde{m}
\]
\end{enumerate}

The algorithm converges to all roots $\omega_j$ of the characteristic equation \eqref{eqn:frequencies_dsm} falling within the selected circle $\Gamma$, including multiple roots \cite{sakurai2003projection,asakura2009numerical,ikegami2010filter}. Circles of increasing radius and centred at the origin can be considered to explore the complex plane and calculate all the eigenvalues requested for practical purposes.

The choice of the parameters $K$, $L_0$, $N_0$ determines the method accuracy. As suggested in ref.~\citep{sakurai2013efficient}, the maximum moment degree $K$ can be set equal to $N_0/4$, in order to preserve both computational cost and numerical accuracy; the minimum number of source vectors $L_0$ is such that $\sigma_{min}/\sigma_{1} < \epsilon$ with small $\epsilon > 0$; the number of quadrature points $N_0$ determines the quadrature error and can be fixed in advance. 
{\sloppy
\subsection{Complex modal analysis}\label{sec:complex_modal}
Now, eigenvalues and eigenfuctions calculated from Eq.~\eqref{eqn:frequencies_dsm} will be used to derive exact analytical expressions for modal impulse and frequency response functions.
The first step is the derivation of proper orthogonality conditions. Eq.~\eqref{eqn:disp_freq}-\eqref{eqn:rot_freq} for the $n^{\mathrm{th}}$ mode without external loads are:
\begin{small}
\begin{flalign}
&GA\left(\gmtder{2}{V_n}{x}+\gtder{\Phi_n}{x}\right)+\rho A \omega_{n}^2 V_n - \sum_{j=1}^{N}k_{eq}(\omega_n)V_{n}(x_j) \delta(x-x_j) = 0 && \label{eqn:disp_freq_eigen}\\
&EI\gmtder{2}{\Phi_n}{x}- GA\left(\gtder{V_n}{x}+\Phi_n\right)+\rho I \omega_{n}^2 \Phi_n = 0 && \label{eqn:rot_freq_eigen}
\end{flalign}
\end{small}
}

\noindent
Multiplying Eq.~\eqref{eqn:disp_freq_eigen} by $V_m$ and Eq.~\eqref{eqn:rot_freq_eigen} by $\Phi_m$, summing the two equations and integrating over $[0,L]$ yield
\begin{small}
\begin{flalign}\label{eqn:mult_n}
&\int_{0}^{L}GA\gtder{V_n}{x}\gtder{V_m}{x}~\text{d}x+ \int_{0}^{L}EI \gtder{\Phi_n}{x}\gtder{\Phi_m}{x}~\text{d}x + \int_{0}^{L}GA\Phi_n\gtder{V_m}{x}~\text{d}{x}  && \nonumber\\
&+ \int_{0}^{L}GA\gtder{V_n}{x}\Phi_m~\text{d}{x} + \int_{0}^{L}GA\Phi_n\Phi_m~\text{d}x + \mathcal{O}_1(\omega_n) =0  &&
\end{flalign}
\end{small}
where:
\begin{small}
\begin{flalign}\label{eqn:O1}
\mathcal{O}_1 (\omega_n) &=  \sum_{j=1}^{N}k_{eq}(\omega_n)V_{n}(x_j)V_{m}(x_j) - \omega^2_{n} \left(\rho A \int_{0}^{L}V_{m}V_{n}~\text{d}x\right. &&\\ 
&\left.+ \rho I \int_{0}^{L}\Phi_m \Phi_{n}~\text{d}x \right)&&\nonumber
\end{flalign}
\end{small}

\noindent
Eq.~\eqref{eqn:mult_n} is obtained integrating by parts, assuming homogeneous B.C. for the beam.

Likewise, multiplying Eq.~\eqref{eqn:disp_freq_eigen} and Eq.~\eqref{eqn:rot_freq_eigen} for the $m^{\mathrm{th}}$ mode by $V_n$ and $\Phi_n$, respectively, summing the two equations and integrating over $[0,L]$ leads to the following equation:
\begin{small}
\begin{flalign}\label{eqn:mult_m}
&\int_{0}^{L}GA\gtder{V_m}{x}\gtder{V_n}{x}~\text{d}x+ \int_{0}^{L}EI \gtder{\Phi_m}{x}\gtder{\Phi_n}{x}~\text{d}x +\int_{0}^{L}GA\Phi_m\gtder{V_n}{x}~\text{d}{x} \nonumber &&\\
&+ \int_{0}^{L}GA\gtder{V_m}{x}\Phi_n~\text{d}{x}+ \int_{0}^{L}GA\Phi_m\Phi_n~\text{d}x +\mathcal{O}_1(\omega_m)
\end{flalign}
\end{small}

\noindent
where $\mathcal{O}_1(\omega_m)$ is Eq.~\eqref{eqn:O1} evaluated for $\omega_m$.
The difference between Eq.~\eqref{eqn:mult_n} and Eq.~\eqref{eqn:mult_m} yields the first orthogonality condition:
\begin{small}
\begin{flalign}\label{eqn:first_orthogonality}
&(k_{eq}(\omega_n)-k_{eq}(\omega_m))\sum_{j=1}^{N}V_{n}(x_j)V_{m}(x_j)\\ \nonumber 
&+ (\omega_{m}^2 -\omega_{n}^2) \left(\rho A \int_{0}^{L}V_nV_{m}~\text{d}x + \rho I\int_{0}^{L}\Phi_n \Phi_{m}~\text{d}x\right) = 0&&
\end{flalign}
\end{small}
Next, the difference between Eq.~\eqref{eqn:mult_n} multiplied by $\omega_m$ and Eq.~\eqref{eqn:mult_m} multiplied by $\omega_n$ provides the second orthogonality condition:
\begin{small}
\begin{flalign}\label{eqn:second:orthogonality}
&(\omega_m-\omega_n)\int_{0}^{L}\left[GA\left(\gtder{V_n}{x}\gtder{V_m}{x} + \Phi_n\gtder{V_m}{x} + \gtder{V_n}{x}\Phi_m + \Phi_n\Phi_m\right)\right. \nonumber  &&\\ 
&\left.+ EI \gtder{\Phi_n}{x}\gtder{\Phi_m}{x}\right]\text{d}x + \sum_{j=1}^{N}(\omega_m k_{eq}(\omega_n)-\omega_n k_{eq}(\omega_m))V_{n}(x_j)V_{m}(x_j)+ \nonumber&&\\ 
&\omega_n \omega_m (\omega_m - \omega_n)\left(\rho A \int_{0}^{L}V_nV_{m}~\text{d}x + \rho I\int_{0}^{L}\Phi_n \Phi_{m}~\text{d}x\right) = 0&&
\end{flalign}
\end{small} 
The orthogonality conditions \eqref{eqn:first_orthogonality}-\eqref{eqn:second:orthogonality} are the basis to derive the modal response, as explained below.

Be the beam subjected to an impulsive loading $p_v(x,t)=f_v(x)\delta(t)$ and $p_\phi(x,t)=f_\phi(x)\delta(t)$, where $\delta(t)$ is a Dirac's delta in time and $f(x)$ a space-dependent function. Adopting the approach in ref.~\cite{failla2020exact,adam2017moving}, the vector  of the beam impulse response functions (IRFs) can be represented by modal superposition as
\begin{equation}\label{eqn:modal_time}
\vect{h}(x,t) = \sum_{k=1}^{\infty}\vect{h}_k (x,t) = \sum_{k=1}^{\infty}g_k (t)\vect{Y}_k(x)
\end{equation}
\begin{equation}\label{eqn:modal_time_coefficient}
g_{k}(t) = \hat{g}_{k}\mathrm{e}^{\mathrm{i}\omega_k t}
\end{equation}
being $\hat{g}_{k}$ a complex coefficient, while $\omega_k$ and $\vect{Y}_{k}(x)$ are eigenvalue and vector of eigenfunctions associated with the $k^{\mathrm{th}}$ mode. Namely, $\omega_k$ and $\vect{Y}_{k}(x)$ are complex as damping of the locally-resonant sandwich beam in Figure~\ref{fig:figure1} is, in general, not proportional. 

Now, replace Eq.~\eqref{eqn:modal_time} for $v(x,t)$ and $\phi(x,t)$ in Eqs.~\eqref{eqn:const_1}-\eqref{eqn:const_2} and multiply Eq.~\eqref{eqn:const_1} by the $n^{\mathrm{th}}$ eigenfunction $V_n(x)$ and Eq.~\eqref{eqn:const_2} by the $n^{\mathrm{th}}$ rotation eigenfunction $\Phi_n(x)$, integrate over $[0,L]$ and sum up the two equations; next, use the two orthogonality conditions \eqref{eqn:first_orthogonality}-\eqref{eqn:second:orthogonality} to decouple the equations in the unknown complex functions  and integrate over $[0^{-},0^{+}]$ obtaining the following expression for every coefficient $\hat{g}_{k}$ :
\begin{small}
\begin{flalign}
\hat{g}_{k} &= \chi_k(\mathrm{i}\omega_k \Pi_{k})^{-1}\label{eqn:ghat}\\
\chi_{k} &= \int_{0}^{L} f_v(x) V_{k}~\mathrm{d}x + \int_{0}^{L}f_\phi(x) \Phi_{k}(x)~\mathrm{d}x\label{eqn:chi}\\
\label{eqn:pi} \Pi_{k} &= \sum_{j=1}^{N}\omega_{k}^{-2}\mu_{j}(\omega_k)V_{k}^{2}(x_j) + 2\rho A \int_{0}^{L}V_{k}^{2}(x)~\text{d}x\\ \nonumber
&+ 2\rho I \int_{0}^{L}\Phi_{k}^{2}(x)~\text{d}x && 
\end{flalign}
\end{small}
where:
\begin{equation}\label{eqn:mu}
\mu_{j}(\omega_k) = \mu(\omega_k) = \lim_{\omega_{n}\rightarrow{\omega_k}}\frac{\omega_n (k_{eq}(\omega_n) -k_{eq}(\omega_k))}{\omega_k - \omega_n}
\end{equation}
The limit \eqref{eqn:mu} can be calculated in analytical form for typical resonators starting from the pertinent frequency-dependent stiffness \cite{failla2020exact,adam2017moving} and examples will be given for the applications in Section 5.

Now, for damping levels typical of engineering applications, the modes contributing to the beam response occur in complex conjugate pairs, i.e. $g_{k}(t)$ in Eq.~\eqref{eqn:modal_time_coefficient} may be $g_{k}(t) = \hat{g}_{k}\mathrm{e}^{\mathrm{i}\omega_k t}$ as well as $g_{k}(t) = \hat{g}_{k}^{*}\mathrm{e}^{-\mathrm{i}\omega_k^{*} t}$ where (*) denotes complex conjugate. The result is the following real form for the modal IRFs of the $k^{\mathrm{th}}$ mode in Eq.~\eqref{eqn:modal_time} \cite{oliveto1997complex}:
\begin{equation}\label{eqn:modal_IRF_contr}
\vect{h}_k(x,t) = \vecg{\gamma}_k(x)|\omega_k|w_k(t) + \vecg{\psi}_{k}(x)\dot{w}_{k}(t)
\end{equation}
where:
\begin{flalign}
&\vecg{\gamma}_{k}(x) = \xi_k \vecg{\psi}_{k}(x) - \sqrt{1-\xi_k^2}\vecg{\upsilon}_k(x) \label{eqn:gamma}\\
&\vecg{\psi}_{k}(x) = 2\operatorname{Re}[\hat{g}_{k}\vect{Y}_{k}(x)] \quad \vecg{\upsilon}_{k}(x) = 2\operatorname{Im}[\hat{g}_{k}\vect{Y}_{k}(x)]\\
&w_{k}(t)=\frac{1}{\omega_{Dk}}\mathrm{e}^{-\xi_k |\omega_k|t}\sin(\omega_{Dk}t); \quad \omega_{Dk}=|\omega_k|\sqrt{1-\xi_{k}^2} &&
\end{flalign}
being $\xi_{k}= \operatorname{Im}(\omega_k)/|\omega_k|$ the modal damping ratio. Based again on ref.~\cite{oliveto1997complex}, the corresponding vector~ 
$\vect{H}_{k} = \Big[
H_{v,k}\\H_{\phi,k}~H_{m,k}~H_{s,k}
\Big]^{\mathrm{T}}$ 
of modal FRFs is
\begin{flalign}\label{eqn:modal_FRF_contr}
\vect{H}_{k}(x,\omega) = \vecg{\gamma}_k(x) |\omega_k|H_{k}(\omega) + \vecg{\psi}_k(x) \mathrm{i}\omega_k H_k(\omega) &&
\end{flalign}
\begin{flalign}\label{eqn:Hk}
H_{k}(\omega) = \frac{1}{|\omega_k|^2 - \omega^2 + \mathrm{i}2\zeta_k|\omega_k|\omega_k}&&
\end{flalign}
Using Eqs.~\eqref{eqn:modal_IRF_contr}-\eqref{eqn:modal_FRF_contr}, the following approximations of the beam IRF and FRF can be built, providing insight into the single modal contributions:
\begin{equation}\label{eqn:modal_IRF}
\vect{h}(x,t) \approx \sum_{k=1}^{M}\vect{h}_{k}(x,t)
\end{equation}
\begin{equation}\label{eqn:modal_FRF}
\vect{H}(x,\omega) \approx \sum_{k=1}^{M}\vect{H}_{k}(x,\omega)
\end{equation}
where $M$ is the number of modes retained for practical applications. Eq.~\eqref{eqn:modal_IRF} and Eq.~\eqref{eqn:modal_FRF} hold for any number of resonators along the beam. Every modal contribution \eqref{eqn:modal_IRF_contr} and \eqref{eqn:modal_FRF_contr} is exact and readily obtainable in analytical form once the eigenvalues are calculated. For practical purposes, a sufficient number of modes $M$ shall be retained in Eq.~\eqref{eqn:modal_IRF} and Eq.~\eqref{eqn:modal_FRF} to obtain approximate yet accurate expressions of IRF and FRF. The IRF and FRF in every resonator follow from Eq.~\eqref{eqn:modal_IRF} and Eq.\eqref{eqn:modal_FRF}, provided that $\vect{Y}_k$ is replaced with $\vect{U}_{j,k}$, i.e. the vector of eigenfunctions associated with the $k^{\mathrm{th}}$ mode for the response in the $j^{\mathrm{th}}$ resonator; $\vect{U}_{j,k}$ can be obtained from the deflection $V_{k}(x_j)$ of the application point.

Finally, a few remarks are in order on the calculation of the transmittance of a cantilever beam within the framework outlined above. Eq.~\eqref{eqn:modal_FRF} calculates the frequency response to any arbitrary load, provided that the eigenfunctions fulfil homogeneous B.C. as, in fact, this is the assumption made when deriving the orthogonality conditions. Moving from this observation, the calculation of the transmittance by Eq.~\eqref{eqn:modal_FRF} can be pursued using the eigenfunctions with homogeneous B.C. and representing the ground displacement $V_{g}\mathrm{e}^{\mathrm{i}\omega t}$ at $x=0$ as a relative deflection between the section at $x=0$ (i.e. fixed) and the section at $x = 0^{+}$. Notice that, in the frequency domain, a relative deflection between adjacent sections at any abscissa $x=x_0$ can be modelled as:
\begin{equation}\label{eqn:disp_discontinuity}
\gtder{V}{x} = \frac{S}{GA} - \Phi + V_{g}\delta(x-x_0) 
\end{equation}
and the corresponding equations of motion are
\begin{footnotesize}
\begin{flalign} 
&GA\left[\gmtder{2}{V}{x}+\gder{\Phi}{x} - V_{g}\delta^{(1)}(x-x_0)\right]+\rho A \omega^2 V + \sum_{j=1}^{N}R_j \delta(x-x_j) = 0 \label{eqn:disp_delta_freq}\\
&EI\gmtder{2}{\Phi}{x}- GA\left[\gder{V}{x}+\Phi- V_{g}\delta(x-x_0)\right]+\rho I \omega^2 \Phi = 0 \label{eqn:rot_delta_freq}&&
\end{flalign}
\end{footnotesize}
In view of Eq.~\eqref{eqn:disp_freq}-\eqref{eqn:rot_freq} and Eq.~\eqref{eqn:chi}, the calculation of the transmittance by Eq.~\eqref{eqn:modal_FRF} involves considering the following term in Eq.~\eqref{eqn:chi}
\begin{small}
\begin{flalign}\label{eqn:chi2}
&\chi_{k} = GAV_g\left[\int_{0}^{L} V_{k}(x)\delta^{(1)}(x-x_0) + \Phi_{k}(x)\delta(x-x_0) ~\text{d}x\right]&& \\ \nonumber
&= GAV_g \left[\Phi_{k}(x_0)-\gtder{V_{k}(x_0)}{x}\right]&&
\end{flalign}
\end{small}
The integral in Eq.~\eqref{eqn:chi2} can be easily solved integrating by parts, providing closed analytical forms for the modal representation \eqref{eqn:modal_FRF} of the transmittance.

{\sloppy 
A final and important remark is that the proposed modal solutions \eqref{eqn:modal_IRF}-\eqref{eqn:modal_FRF} can be easily extended to consider the mass-spring-dashpot subsystems as exerting distributed forces over the mutual distance \cite{chen2011containing,chen2011dynamic}, see Appendix~B for details.

}
\section{Numerical applications}
\begin{figure}[h]
\centering
\includegraphics[scale=0.85]{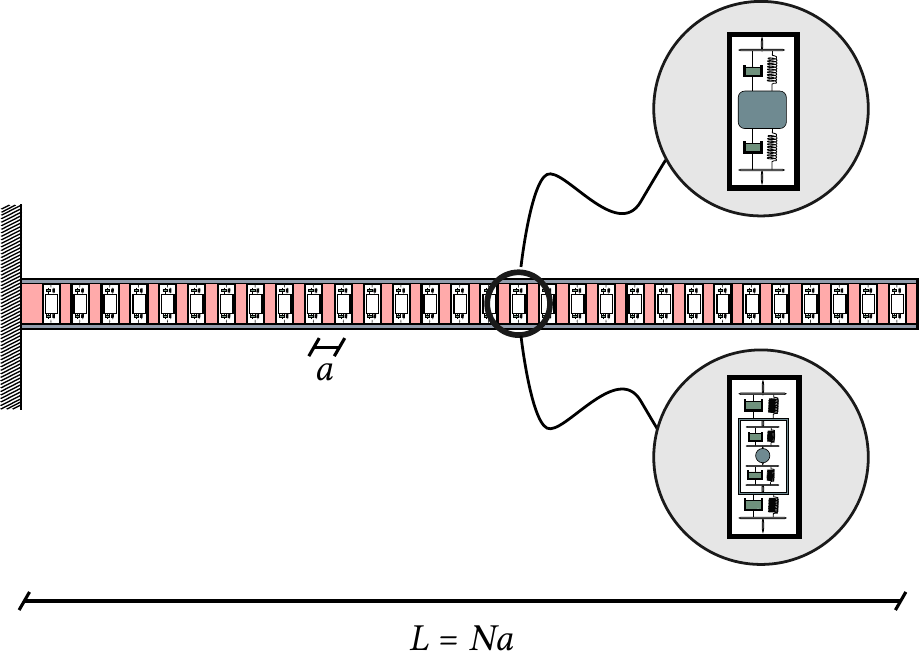}
\caption{\label{fig:figure2}Cantilever locally-resonant sandwich beam hosting 1-DOF or 2-DOF resonators.}
\end{figure}
Consider the cantilever locally-resonant sandwich beam in Figure~\ref{fig:figure2}. Following ref.~\cite{chen2011containing}, parameters (3)-(6) of the equivalent single-layer Timoshenko beam model are: \\ $EI = \SI{611}{\N~\unexp{\m}{-2}}$; $GA = \SI{1.12\times 10^4}{\N}$; $\rho A = \SI{0.1248}{\kg~\unexp{\m}{-1}}$; $\rho I = \SI{1.69\times 10^{-5}}{\kg~\m}$; $a = \SI{0.01}{\m}$ is the mutual distance between the resonators, $N = 30$ is the number of the resonators, $L = \SI{0.30}{\m}$ is the total length of the beam. 

Two cases are considered: (a) 1-DOF resonators with parameters $k_1 = \SI{7415.74}{\N~\unexp{\m}{-1}}$, $c_1 = \SI{0.05}{\N~\s~\unexp{m}{-1}}$, $m_1 = \SI{0.00117}{\kg}$; (b) 2-DOF resonators with parameters $k_1 = k_2 = \SI{13361.97}{\N~\unexp{\m}{-1}}$; $c_1 =  \SI{0.05}{\N~\s~\unexp{m}{-1}}$; $c_2 =  \SI{0.113}{\N~\s~\unexp{m}{-1}}$; $m_1 = \SI{0.0047}{\kg}$; $m_2 = \SI{0.019}{\kg}$. 
The solution methods proposed in Sections 3-4 are applied to both cases. The modal expansions \eqref{eqn:modal_IRF}-\eqref{eqn:modal_FRF} for IRF and FRF require calculating the limit \eqref{eqn:mu} depending on the frequency-dependent stiffness \eqref{eqn:equivalent_DS} pertinent to the 1-DOF and 2-DOF resonators and available in the following forms:
\begin{itemize}
\item[] \textbf{1-DOF}
\begin{small}
\begin{flalign}\label{eqn:mu_1dof}
\mu(\omega_k) = \frac{m_1 \omega _k^2 \left(-2 c_1^2 \omega _k^2- \mathrm{i} c_1 m_1 \omega _k^3+4 \mathrm{i} c_1 k_1 \omega _k+2 k_1^2\right)}{\left(k_1 + \mathrm{i} c_1 \omega _k+m_1 \omega _k^2\right){}^2}&&
\end{flalign}
\end{small}
\item[] \textbf{2-DOF}
\begin{small}
\begin{flalign}\label{eqn:mu_2dof}
\mu(\omega_k) =\frac{\omega_k^{2}(c_1 \Delta _1 \omega_k ^2+2 c_1 \Gamma _2 k_1 \omega_k +\Gamma _1 k_1^2)}{\left(\gamma _1 k_1+\omega_k  \left(\gamma _2 \omega_k +i c_1 \gamma _1\right)\right){}^2}&&
\end{flalign}
\end{small}
where:
\begin{flalign}
&\Gamma _1=4 \mathrm{i} k_2 \omega_k  \left(c_2 \left(m_1+m_2\right)+\mathrm{i} m_1 m_2 \omega_k \right)+\omega_k ^2 \nonumber\\   
&\left(-\mathrm{i} c_2 m_2 \left(4 m_1+m_2\right) \omega_k -2 c_2^2 \left(m_1+m_2\right) \right. \nonumber  &&\\
& \left. +2 m_1 m_2^2 \omega_k ^2\right)  +2 k_2^2 \left(m_1+m_2\right)&&\\ 
&\Gamma _2=-4 k_2 \omega_k  \left(c_2 \left(m_1+m_2\right)+\mathrm{i} m_1 m_2 \omega_k \right) \nonumber  &&\\ 
&+\omega_k ^2 \left(c_2 m_2  \left(4 m_1+m_2\right) \omega_k -2 \mathrm{i} c_2^2 \left(m_1+m_2\right)\right. \nonumber && \\
&\left. +2 \mathrm{i} m_1 m_2^2 \omega_k ^2\right)+ 2 \mathrm{i} k_2^2 \left(m_1+m_2\right)&&\\
&\Delta _1=c_1 \Gamma _3-\mathrm{i} \omega_k  \left(\mathrm{i} c_2 \left(m_1+m_2\right) \omega_k +k_2 \left(m_1+m_2\right) \nonumber \right. &&\\
&\left. -m_1 m_2 \omega_k ^2\right){}^2&&\\
&\Gamma _3=4 k_2 \omega_k  \left(m_1 m_2 \omega_k -\mathrm{i} c_2 \left(m_1+m_2\right)\right)+\omega_k ^2 \left(\mathrm{i} c_2 m_2 \right. \nonumber && \\
&\left. \left(4 m_1+m_2\right) \omega_k +2 c_2^2 \left(m_1+m_2\right)-2 m_1 m_2^2 \omega_k ^2\right) \nonumber &&\\
&-2 k_2^2 \left(m_1+m_2\right)&&\\
&\gamma _1=k_2+\omega_k  \left(-m_2 \omega_k +\mathrm{i} c_2\right)&&\\
&\gamma _2=-k_2 \left(m_1+m_2\right)+\omega_k  \left(m_1 m_2 \omega_k \right. \nonumber &&\\ 
& \left. -\mathrm{i} c_2 \left(m_1+m_2\right)\right)&&
\end{flalign}
\end{itemize}
The proposed solution in and the contour-integral algorithm in Sec.~\ref{sec:fr_response}-\ref{sec:modal_analysis} are implemented in Matlab \citep{MATLAB:2010}.
\subsection{1-DOF resonators}
\begin{figure}[h]
\centering
\includegraphics[scale=0.545]{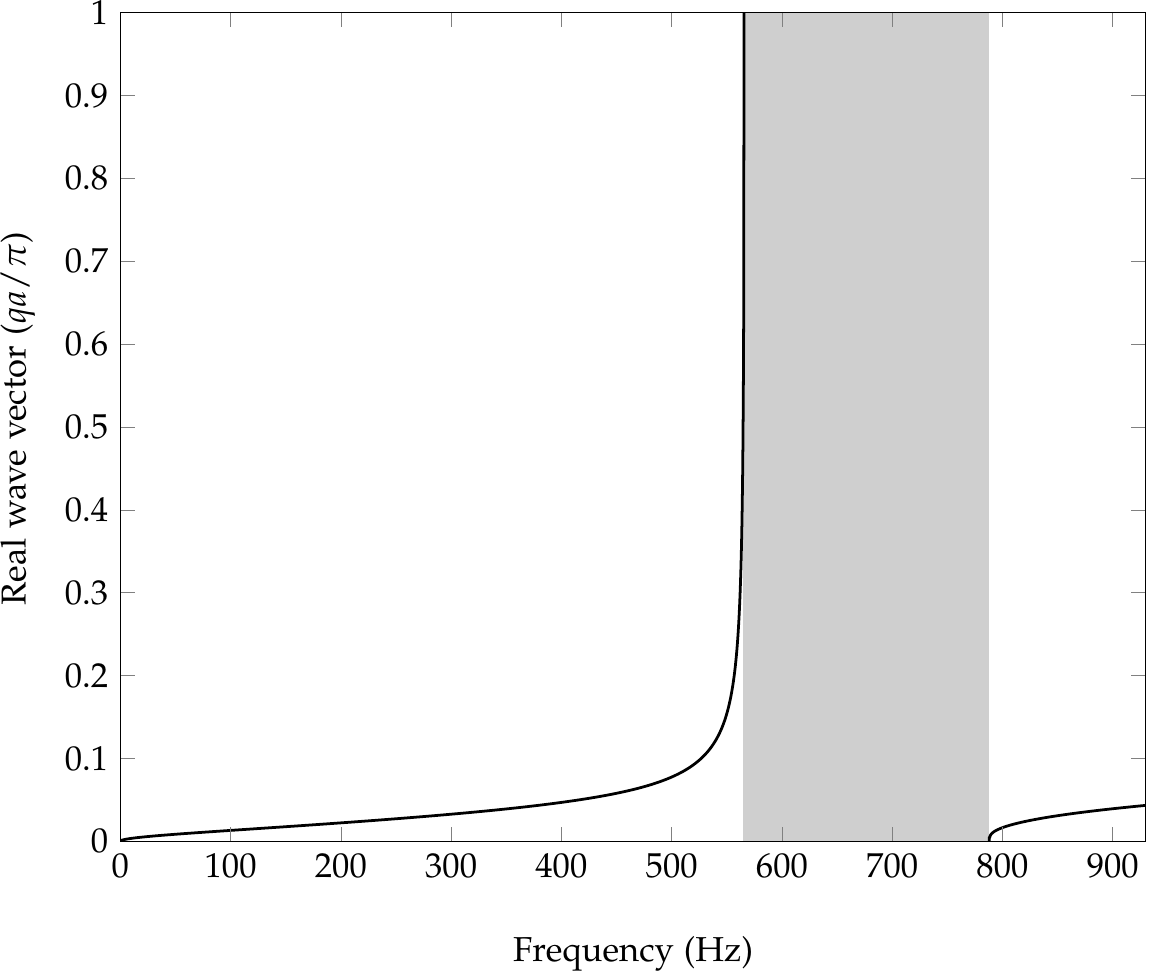}
\caption{\label{fig:figure3}Band gaps of the infinite locally-resonant sandwich beam in Figure~\ref{fig:figure2} with 1-DOF resonators.}
\end{figure}

For a first insight into the dynamics of the locally-resonant sandwich beam with 1-DOF resonators, the band gaps of the infinite beam with no damping are calculated using a standard transfer matrix approach \cite{liu2007design}. As expected given the fact that every resonator has one DOF, Figure \ref{fig:figure3} shows one band gap, where no real wave vectors are found. The band gap spans the frequency range 565-788~Hz.

Next, attention is focused on the cantilever beam and damping is considered within the resonators. The contour-integral algorithm in Section 4.1 is applied to calculate the first 131 complex eigenvalues, reported in Table \ref{tab:eigen_1dof}. 
\begin{table*}
\resizebox{\textwidth}{!}{
\begin{tabular}{cl}
\toprule
Mode&Eigenvalue\\
\midrule
$1$ &$ 967.680\pm0.062\mathrm{i}$\\ 
$2$ &$ 2498.070\pm4.094\mathrm{i}$\\ 
$3$ &$ 3177.911\pm13.012\mathrm{i}$\\ 
$4$ &$ 3367.718\pm16.915\mathrm{i}$\\ 
$5$ &$ 3445.411\pm18.664\mathrm{i}$\\ 
$6$ &$ 3483.342\pm19.543\mathrm{i}$\\ 
$7$ &$ 3505.011\pm20.051\mathrm{i}$\\ 
$8$ &$ 3518.398\pm20.368\mathrm{i}$\\ 
$9$ &$ 3527.317\pm20.579\mathrm{i}$\\ 
$10$ &$ 3533.517\pm20.727\mathrm{i}$\\ 
$11$ &$ 3538.019\pm20.834\mathrm{i}$\\ 
$12$ &$ 3541.376\pm20.914\mathrm{i}$\\ 
$13$ &$ 3543.950\pm20.975\mathrm{i}$\\ 
$14$ &$ 3545.959\pm21.023\mathrm{i}$\\ 
$15$ &$ 3547.558\pm21.061\mathrm{i}$\\ 
$16$ &$ 3548.845\pm21.092\mathrm{i}$\\ 
$17$ &$ 3549.894\pm21.117\mathrm{i}$\\ 
$18$ &$ 3550.757\pm21.138\mathrm{i}$\\ 
$19$ &$ 3551.472\pm21.155\mathrm{i}$\\ 
$20$ &$ 3552.068\pm21.169\mathrm{i}$\\ 
$21$ &$ 3552.565\pm21.181\mathrm{i}$\\ 
$22$ &$ 3552.981\pm21.191\mathrm{i}$\\ 
$23$ &$ 3553.329\pm21.200\mathrm{i}$\\ 
$24$ &$ 3553.617\pm21.206\mathrm{i}$\\ 
$25$ &$ 3553.854\pm21.212\mathrm{i}$\\ 
$26$ &$ 3554.046\pm21.217\mathrm{i}$\\ 
$27$ &$ 3554.198\pm21.220\mathrm{i}$\\ 
$28$ &$ 3554.312\pm21.223\mathrm{i}$\\ 
$29$ &$ 3554.439\pm21.226\mathrm{i}$\\ 
$30$ &$ 3554.392\pm21.225\mathrm{i}$\\ 
$31$ &$ 5094.551\pm41.909\mathrm{i}$\\ 
$32$ &$ 6076.198\pm37.949\mathrm{i}$\\ 
$33$ &$ 8561.842\pm28.981\mathrm{i}$\\ 
\bottomrule
\end{tabular}
\begin{tabular}{cl}
\toprule
Mode&Eigenvalue\\
\midrule$34$ &$ 11411.589\pm25.056\mathrm{i}$\\ 
$35$ &$ 14453.851\pm23.306\mathrm{i}$\\ 
$36$ &$ 17504.898\pm22.413\mathrm{i}$\\ 
$37$ &$ 20607.405\pm21.905\mathrm{i}$\\ 
$38$ &$ 23699.402\pm21.572\mathrm{i}$\\ 
$39$ &$ 26817.633\pm21.357\mathrm{i}$\\ 
$40$ &$ 29920.903\pm21.175\mathrm{i}$\\ 
$41$ &$ 33041.580\pm21.038\mathrm{i}$\\ 
$42$ &$ 36133.097\pm20.791\mathrm{i}$\\ 
$43$ &$ 39207.301\pm20.081\mathrm{i}$\\ 
$44$ &$ 41615.636\pm8.516\mathrm{i}$\\ 
$45$ &$ 42883.789\pm14.665\mathrm{i}$\\ 
$46$ &$ 45683.115\pm20.384\mathrm{i}$\\ 
$47$ &$ 48777.480\pm20.690\mathrm{i}$\\ 
$48$ &$ 51886.117\pm20.764\mathrm{i}$\\ 
$49$ &$ 55012.532\pm20.766\mathrm{i}$\\ 
$50$ &$ 58136.869\pm20.764\mathrm{i}$\\ 
$51$ &$ 61268.083\pm20.751\mathrm{i}$\\ 
$52$ &$ 64397.026\pm20.740\mathrm{i}$\\ 
$53$ &$ 67530.045\pm20.726\mathrm{i}$\\ 
$54$ &$ 70661.089\pm20.712\mathrm{i}$\\ 
$55$ &$ 73795.070\pm20.696\mathrm{i}$\\ 
$56$ &$ 76927.227\pm20.678\mathrm{i}$\\ 
$57$ &$ 80061.793\pm20.656\mathrm{i}$\\ 
$58$ &$ 83194.469\pm20.624\mathrm{i}$\\ 
$59$ &$ 86329.326\pm20.574\mathrm{i}$\\ 
$60$ &$ 89461.808\pm20.463\mathrm{i}$\\ 
$61$ &$ 92595.944\pm19.948\mathrm{i}$\\ 
$62$ &$ 95728.273\pm21.431\mathrm{i}$\\ 
$63$ &$ 98858.127\pm20.857\mathrm{i}$\\ 
$64$ &$ 101922.582\pm18.719\mathrm{i}$\\
$65$ &$ 102800.655\pm2.539\mathrm{i}$\\ 
$66$ &$ 105167.218\pm20.683\mathrm{i}$\\ 
\bottomrule
\end{tabular}
\begin{tabular}{cc}
\toprule
Mode&Eigenvalue\\
\midrule $67$ &$  108291.328\pm20.777\mathrm{i}$\\ 
$68$ &$  111421.313\pm20.780\mathrm{i}$\\ 
$69$ &$  114556.083\pm20.768\mathrm{i}$\\ 
$70$ &$  117690.008\pm20.761\mathrm{i}$\\ 
$71$ &$  120825.715\pm20.752\mathrm{i}$\\ 
$72$ &$  123960.628\pm20.746\mathrm{i}$\\ 
$73$ &$  127096.642\pm20.739\mathrm{i}$\\ 
$74$ &$  130231.990\pm20.734\mathrm{i}$\\ 
$75$ &$  133368.169\pm20.729\mathrm{i}$\\ 
$76$ &$  136503.753\pm20.725\mathrm{i}$\\ 
$77$ &$  139640.041\pm20.720\mathrm{i}$\\ 
$78$ &$  142775.746\pm20.716\mathrm{i}$\\ 
$79$ &$  145912.108\pm20.712\mathrm{i}$\\ 
$80$ &$  149047.825\pm20.708\mathrm{i}$\\ 
$81$ &$  152184.207\pm20.704\mathrm{i}$\\ 
$82$ &$  155319.733\pm20.699\mathrm{i}$\\ 
$83$ &$  158455.939\pm20.693\mathrm{i}$\\ 
$84$ &$  161590.364\pm20.682\mathrm{i}$\\ 
$85$ &$  164723.935\pm20.643\mathrm{i}$\\ 
$86$ &$  167628.273\pm6.541\mathrm{i}$\\ 
$87$ &$  167974.147\pm14.280\mathrm{i}$\\ 
$88$ &$  171010.497\pm20.629\mathrm{i}$\\ 
$89$ &$  174144.441\pm20.647\mathrm{i}$\\ 
$90$ &$  177279.215\pm20.644\mathrm{i}$\\ 
$91$ &$  180415.506\pm20.618\mathrm{i}$\\ 
$92$ &$  183551.381\pm20.559\mathrm{i}$\\ 
$93$ &$  186687.811\pm20.267\mathrm{i}$\\ 
$94$ &$  189824.457\pm21.142\mathrm{i}$\\ 
$95$ &$  192960.973\pm20.849\mathrm{i}$\\ 
$96$ &$  196097.302\pm20.791\mathrm{i}$\\ 
$97$ &$  199233.997\pm20.766\mathrm{i}$\\ 
$98$ &$  202370.416\pm20.749\mathrm{i}$\\ 
$99$ &$  205507.151\pm20.740\mathrm{i}$\\  
\bottomrule
\end{tabular}
\begin{tabular}{cc}
\toprule
Mode&Eigenvalue\\
\midrule $100$ &$  208643.605\pm20.733\mathrm{i}$\\ 
$101$ &$  211780.361\pm20.728\mathrm{i}$\\ 
$102$ &$  214916.799\pm20.723\mathrm{i}$\\ 
$103$ &$  218053.552\pm20.719\mathrm{i}$\\ 
$104$ &$  221189.870\pm20.715\mathrm{i}$\\ 
$105$ &$  224326.517\pm20.712\mathrm{i}$\\ 
$106$ &$  227462.225\pm20.704\mathrm{i}$\\ 
$107$ &$  230597.380\pm20.682\mathrm{i}$\\ 
$108$ &$  233424.810\pm2.526\mathrm{i}$\\ 
$109$ &$  233780.308\pm18.256\mathrm{i}$\\ 
$110$ &$  236878.026\pm20.686\mathrm{i}$\\ 
$111$ &$  240013.600\pm20.697\mathrm{i}$\\ 
$112$ &$  243149.515\pm20.699\mathrm{i}$\\ 
$113$ &$  246286.213\pm20.698\mathrm{i}$\\ 
$114$ &$  249422.685\pm20.696\mathrm{i}$\\ 
$115$ &$  252559.510\pm20.694\mathrm{i}$\\ 
$116$ &$  255696.136\pm20.692\mathrm{i}$\\ 
$117$ &$  258833.001\pm20.689\mathrm{i}$\\ 
$118$ &$  261969.694\pm20.685\mathrm{i}$\\ 
$119$ &$  265106.580\pm20.680\mathrm{i}$\\ 
$120$ &$  268243.305\pm20.674\mathrm{i}$\\ 
$121$ &$  271380.204\pm20.664\mathrm{i}$\\ 
$122$ &$  274516.940\pm20.647\mathrm{i}$\\ 
$123$ &$  277653.845\pm20.609\mathrm{i}$\\ 
$124$ &$  280790.559\pm20.420\mathrm{i}$\\ 
$125$ &$  283927.477\pm20.983\mathrm{i}$\\ 
$126$ &$  287064.115\pm20.794\mathrm{i}$\\ 
$127$ &$  290200.948\pm20.755\mathrm{i}$\\ 
$128$ &$  293337.228\pm20.735\mathrm{i}$\\ 
$129$ &$  296473.209\pm20.713\mathrm{i}$\\ 
$130$ &$  299408.258\pm3.362\mathrm{i}$\\ 
$131$ &$  299651.794\pm17.409\mathrm{i}$\\
\\ 
\bottomrule
\end{tabular}
}
\caption{\label{tab:eigen_1dof}Complex eigenvalues of the cantilever locally-resonant sandwich beam in Figure~\ref{fig:figure2} with 1-DOF resonators.
}
\end{table*}
Several eigenvalues are close to each other, as a result of local resonance; remarkably, the algorithm proves capable of capturing also those differing by a few digits.
\begin{figure}[h]
\centering
\includegraphics[scale=0.545]{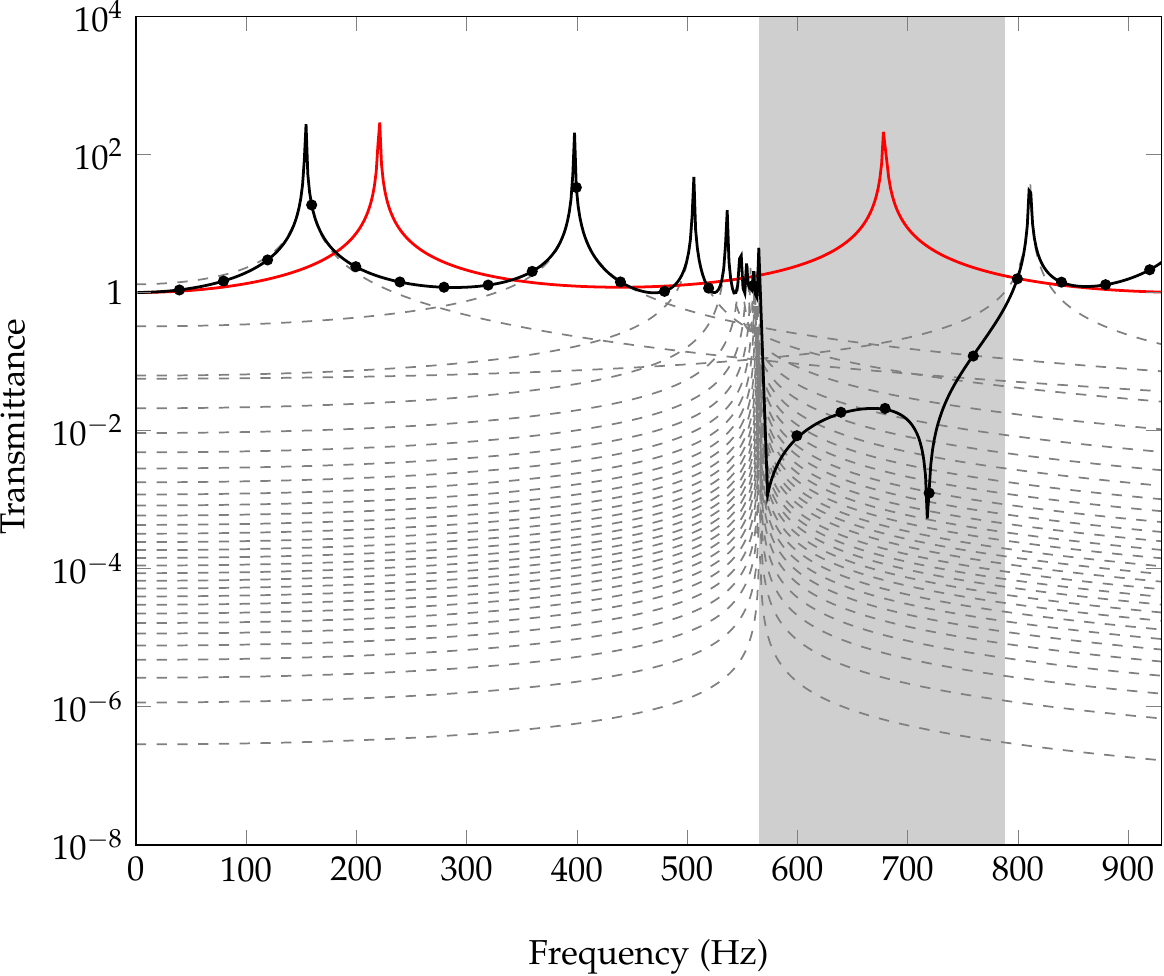}
\caption{\label{fig:trs_1dof}Transmittance of the cantilever locally-resonant sandwich beam in Figure 2 with 1-DOF resonators: exact response \eqref{eqn:exact_FRF} (black continuous line); total modal response \eqref{eqn:modal_FRF} for $M = 131$ (black dots); single modal responses \eqref{eqn:modal_FRF_contr} (gray dashed lines) for $k = 1,...,31$; exact response without resonators (red continuous line).}
\end{figure}
Figure \ref{fig:trs_1dof} shows the transmittance of the cantilever locally-resonant sandwich beam, as calculated using the exact frequency response \eqref{eqn:exact_FRF} with conditions \eqref{eqn:reaction_trasmittance}-\eqref{eqn:BC_trasmittance} and the corresponding modal representation \eqref{eqn:modal_FRF}  including $M=131$ modes, where the coefficients $\chi_k$ are given by Eq.~\eqref{eqn:chi2}; 
\begin{figure}[t]
\captionsetup[subfigure]{labelformat=empty,justification=centering}
\begin{subfigure}{\linewidth}
\centering
\includegraphics[scale=0.545]{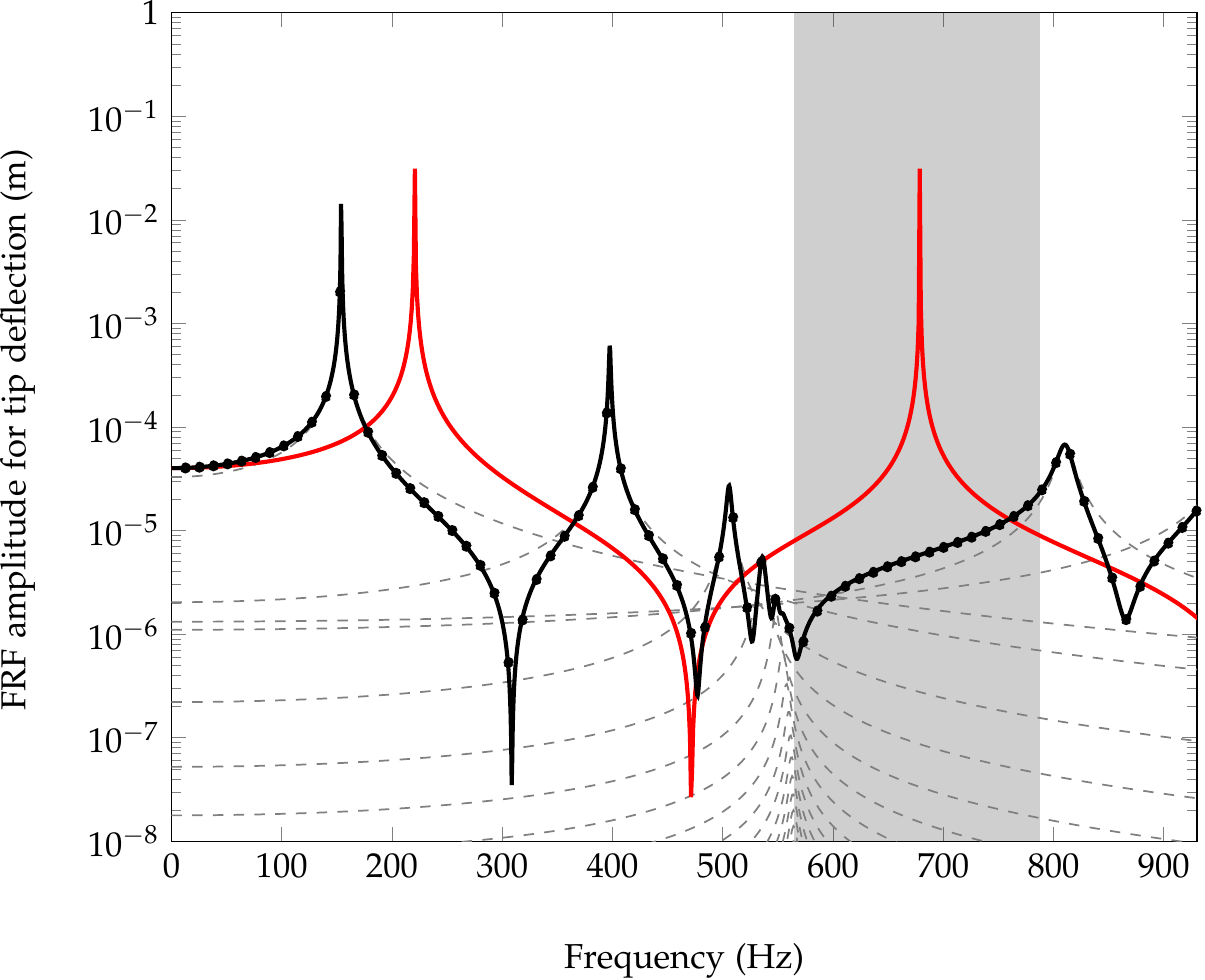}
\caption{\label{fig:frf_1dof_all}~~~~~~~~~~~~~~~~(a)}
\end{subfigure}
\begin{subfigure}{\linewidth}
\centering
\includegraphics[scale=0.545]{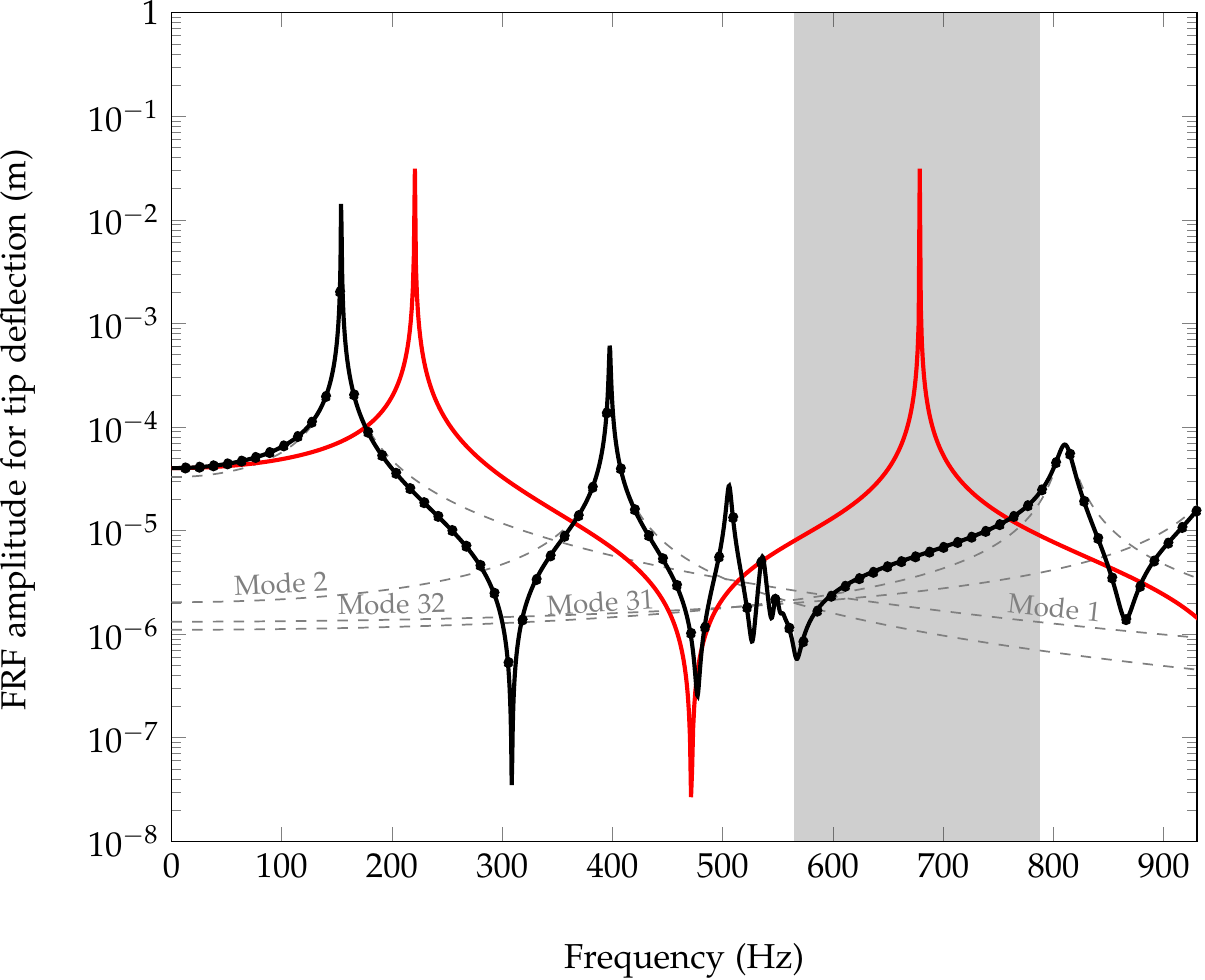}
\caption{\label{fig:frf_1dof_contr}~~~~~~~~~~~~~~~~(b)}
\end{subfigure}
\caption{\label{fig:frf_1dof}FRF for tip deflection of cantilever locally-resonant sandwich beam in Figure 2 with 1-DOF resonators, under a unit harmonic force applied at the free end: exact response \eqref{eqn:exact_FRF} (black continuous line); total modal response \eqref{eqn:modal_FRF} with $M = 131$ (black dots); single modal responses \eqref{eqn:modal_FRF_contr} (gray dashed lines); exact response without resonators (red continuous line); modal responses \eqref{eqn:modal_FRF_contr} are reported for $k=1,...,20$ and $k =30,31,32$ (Fig.~\ref{fig:frf_1dof_all}) and $k=1,2, 31, 32$ (Fig.~\ref{fig:frf_1dof_contr}).}
\end{figure}
additionally, the individual modal contributions \eqref{eqn:modal_FRF_contr} are reported for $k=1,2,...,31$, while the remaining ones up to $M=131$ are omitted for clarity. The two solutions \eqref{eqn:exact_FRF} and \eqref{eqn:modal_FRF} are in perfect agreement, substantiating the correctness of the two approaches proposed in this paper. The transmittance within the band gap is well lower than the transmittance over the remaining frequency domain, meaning that the wave attenuation properties of the infinite beam (see Figure \ref{fig:figure3}) hold also for the finite beam. A further interesting observation is that the peaks of all individual modal contributions occur either below or above the band gap, i.e. there are no resonance modes within the band gap. For completeness, Figure~\ref{fig:trs_1dof} reports the transmittance of the beam without resonators, which exhibits a peak within the band gap well larger than the transmittance of the beam with resonators.

Now, the interest is to calculate the FRF of the cantilever beam acted upon by a unit harmonic force applied at the free end. Figure~\ref{fig:frf_1dof_all} illustrates the FRF for the tip deflection over the frequency range 0-930~Hz, as computed using the exact solution \eqref{eqn:exact_FRF} and the modal representation \eqref{eqn:modal_FRF} for $M=131$; again, the individual modal contributions \eqref{eqn:modal_FRF_contr} are reported for $k=1,2,...,20$ and $k=30,31,32$. The two solutions are in perfect agreement; for the frequency range  considered in Figure~\ref{fig:frf_1dof}, $M=131$ modes are sufficient to provide a very accurate modal representation \eqref{eqn:modal_FRF} of the exact solution \eqref{eqn:exact_FRF}.

Figure \ref{fig:frf_1dof_all} shows also the FRF of the beam without resonators, showing that is generally larger than the FRF of the beam with resonators within the whole band gap, except for a limited frequency range 759-788~Hz, i.e. at the right end of the band gap. The inspection of the modal contributions suggests that this is essentially attributable to the contributions of modes $1-2-31-32$, as highlighted in Figure~\ref{fig:frf_1dof_contr}. 

\begin{figure}[h]
\captionsetup[subfigure]{labelformat=empty,justification=centering}
\begin{subfigure}{\linewidth}
\centering
\includegraphics[scale=0.539]{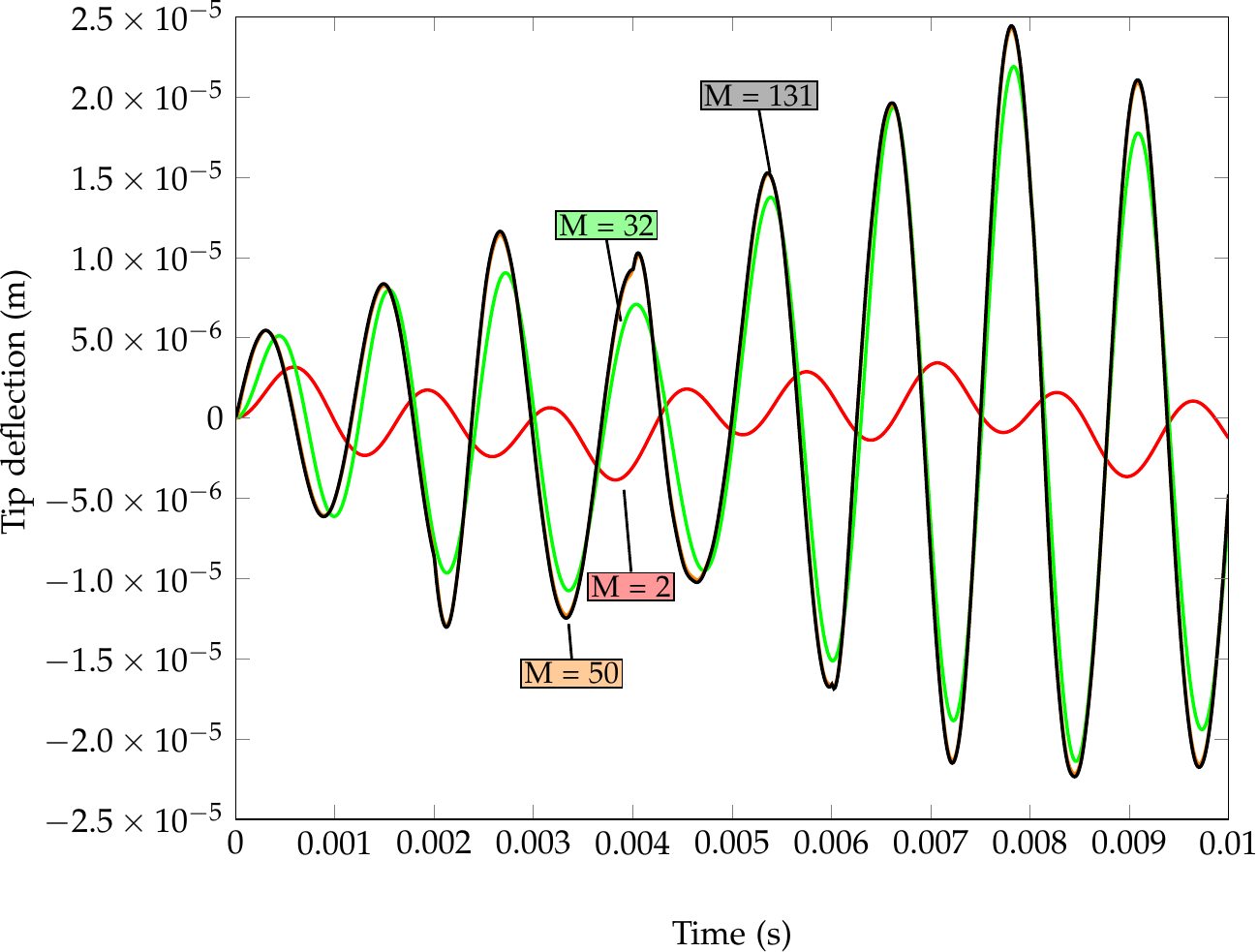}
\caption{\label{fig:irf_1dof_a}~~~~~~~~~~~~~~~~(a)}
\end{subfigure}
\begin{subfigure}{\linewidth}
\centering
\includegraphics[scale=0.539]{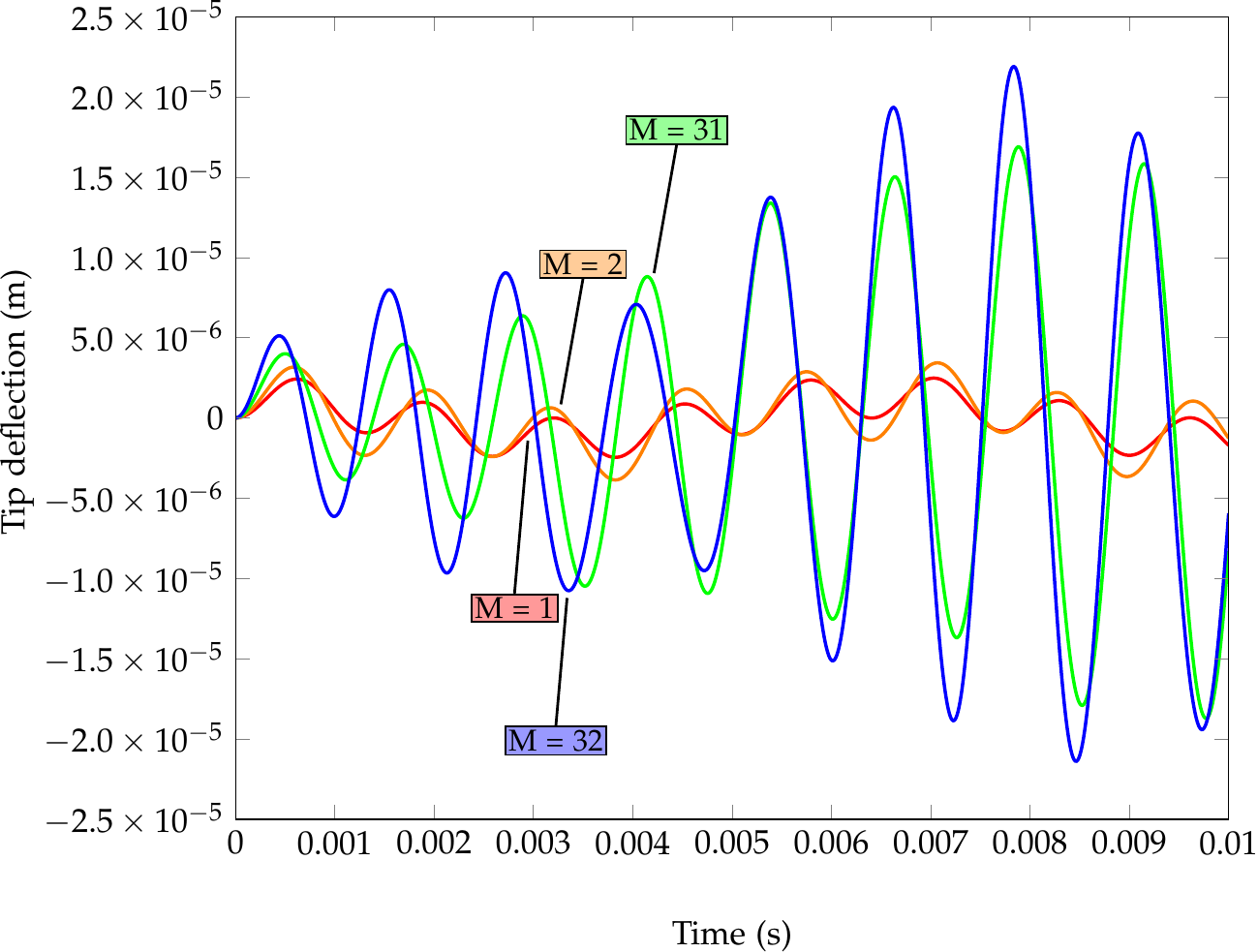}
\caption{\label{fig:irf_1dof_b}~~~~~~~~~~~~~~~~(b)}
\end{subfigure}
\begin{subfigure}{\linewidth}
\centering
\includegraphics[scale=0.539]{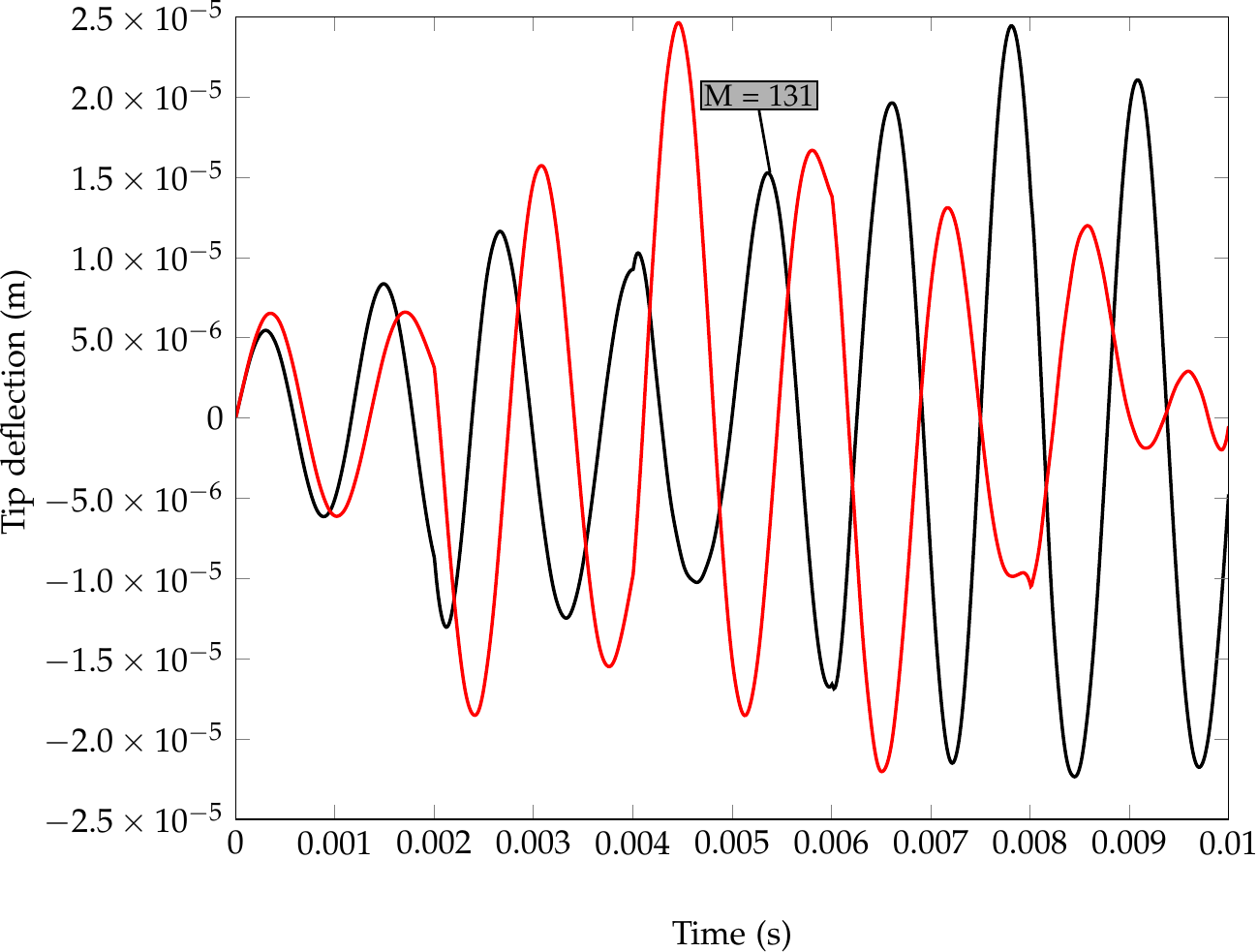}
\caption{\label{fig:irf_1dof_c}~~~~~~~~~~~~~~~~(c)}
\end{subfigure}
\caption{\label{fig:irf_1dof} Tip deflection of the cantilever locally-resonant sandwich beam in Figure~\ref{fig:figure2} with 1-DOF resonators, under a unit cosine force with frequency 780 Hz: (a) total response for increasing number of modes $M$ in Eq.~\eqref{eqn:modal_IRF}; (b) single modal response for most significant modes; (c) total response for $M=131$ in Eq.~\eqref{eqn:modal_IRF} (black continuous line) and response of the beam without resonators (red continuous line).}
\end{figure}

For a further insight into this issue, the time response is investigated. Specifically, the closed analytical expression \eqref{eqn:modal_IRF} for the IRF is used to calculate the tip deflection of the beam acted upon by a unit cosine force with frequency \SI{780}{\Hz}, applied at the free end. Figure \ref{fig:irf_1dof_a} shows no significant changes in the response if more than $M = 50$ modes are included in Eq.~\eqref{eqn:modal_IRF} for the IRF. Further, consistently with the FRF in Figure~\ref{fig:frf_1dof}, Figure~\ref{fig:irf_1dof_b} shows that the most significant contributions to the response are associated with the $1^{\mathrm{st}}$, $2^{\mathrm{nd}}$, $31^{\mathrm{st}}$, $32^{\mathrm{nd}}$ modes; indeed, due mainly to these contributions, the response of the beam with resonators attains almost the same order of magnitude of the response of the beam without resonators, as shown in Figure \ref{fig:irf_1dof_c}. Notice that the insight gained into the modal contributions is a crucial information for design purposes because, e.g., once mass and stiffness of the resonators are calibrated, the damping coefficients might be selected so as to minimize the most significant modal contributions, i.e., in this case, those associated with $1^{\mathrm{st}}$, $2^{\mathrm{nd}}$, $31^{\mathrm{st}}$, $32^{\mathrm{nd}}$ modes. 

This substantiates the interest in the proposed modal representation of the response, in both frequency and time domain.

\begin{figure}[t]
\centering
\includegraphics[scale=0.545]{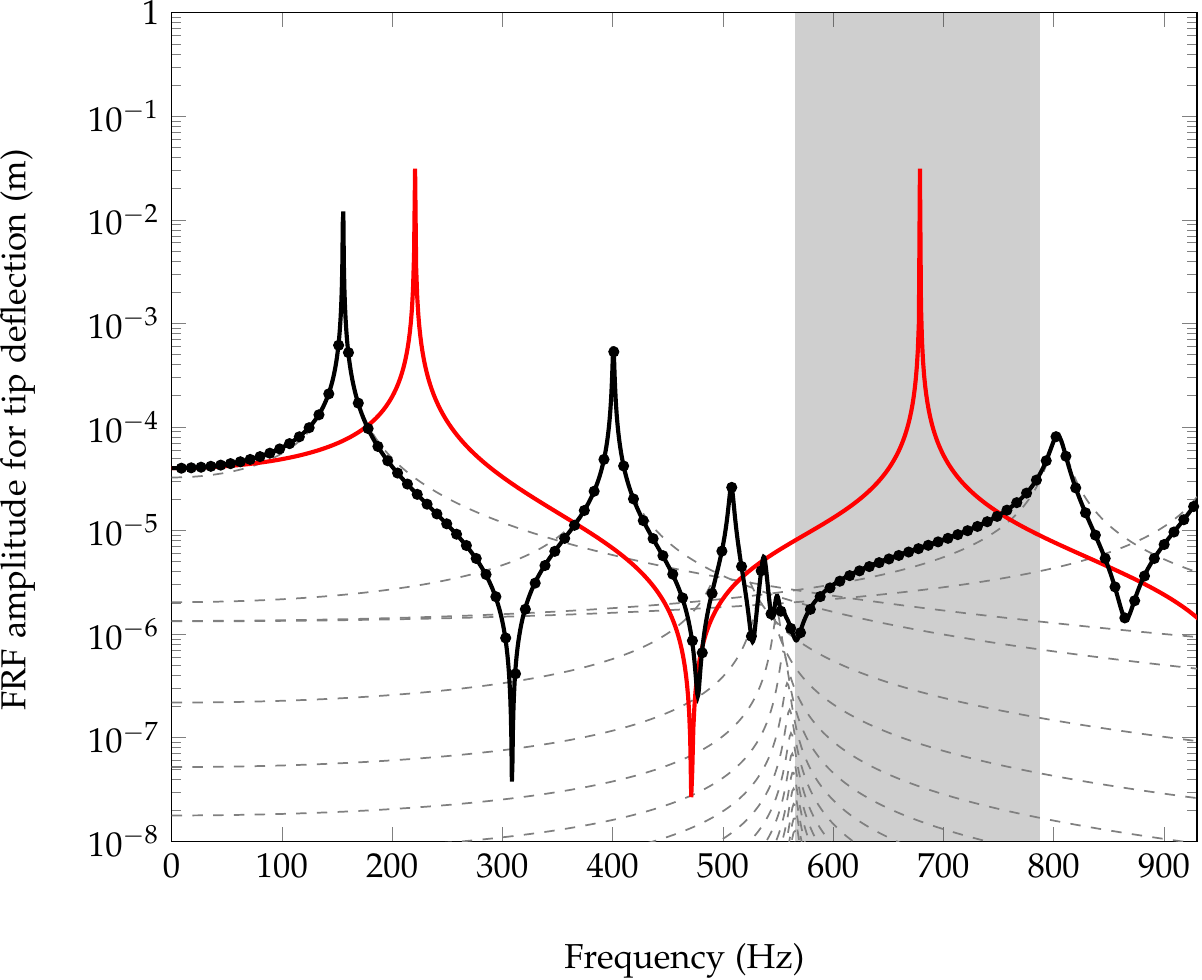}
\caption{\label{fig:frf_1dof_distr} FRF for tip deflection of cantilever locally-resonant sandwich beam in Figure~\ref{fig:figure2} with 1-DOF resonators modelled as exerting distributed forces over the mutual distance $a$, under a unit harmonic force applied at the free end: exact response \eqref{eqn:exact_FRF} (black continuous line); total modal response \eqref{eqn:modal_FRF} for $M = 131$ (black dots); single modal responses \eqref{eqn:modal_FRF_contr} (gray dashed lines); exact response without resonators (red continuous line); modal responses \eqref{eqn:modal_FRF_contr} are reported for $k=1,...,20$ and $k=30,31,32$.}
\end{figure}
Finally, for completeness the proposed solutions are implemented considering the resonators as exerting distributed forces over the mutual distance $a = \SI{0.01}{\m}$. This model can readily be handled with little modifications to Eq.~\eqref{eqn:exact_FRF} for the exact FRF and Eqs.~\eqref{eqn:modal_IRF}-\eqref{eqn:modal_FRF} for modal IRF and FRF, as explained in Appendix~B. The FRF for the tip deflection under a unit harmonic force at the free end in Figure \ref{fig:frf_1dof_distr} are very similar to the corresponding ones reported in Figures \ref{fig:frf_1dof}, in agreement with previous findings in ref.~\cite{chen2011dynamic} for locally-resonant sandwich beams. The same comments hold for the time response, which is not included for conciseness.

\subsection{2-DOF resonators}

\begin{figure}[h]
\centering
\includegraphics[scale=0.545]{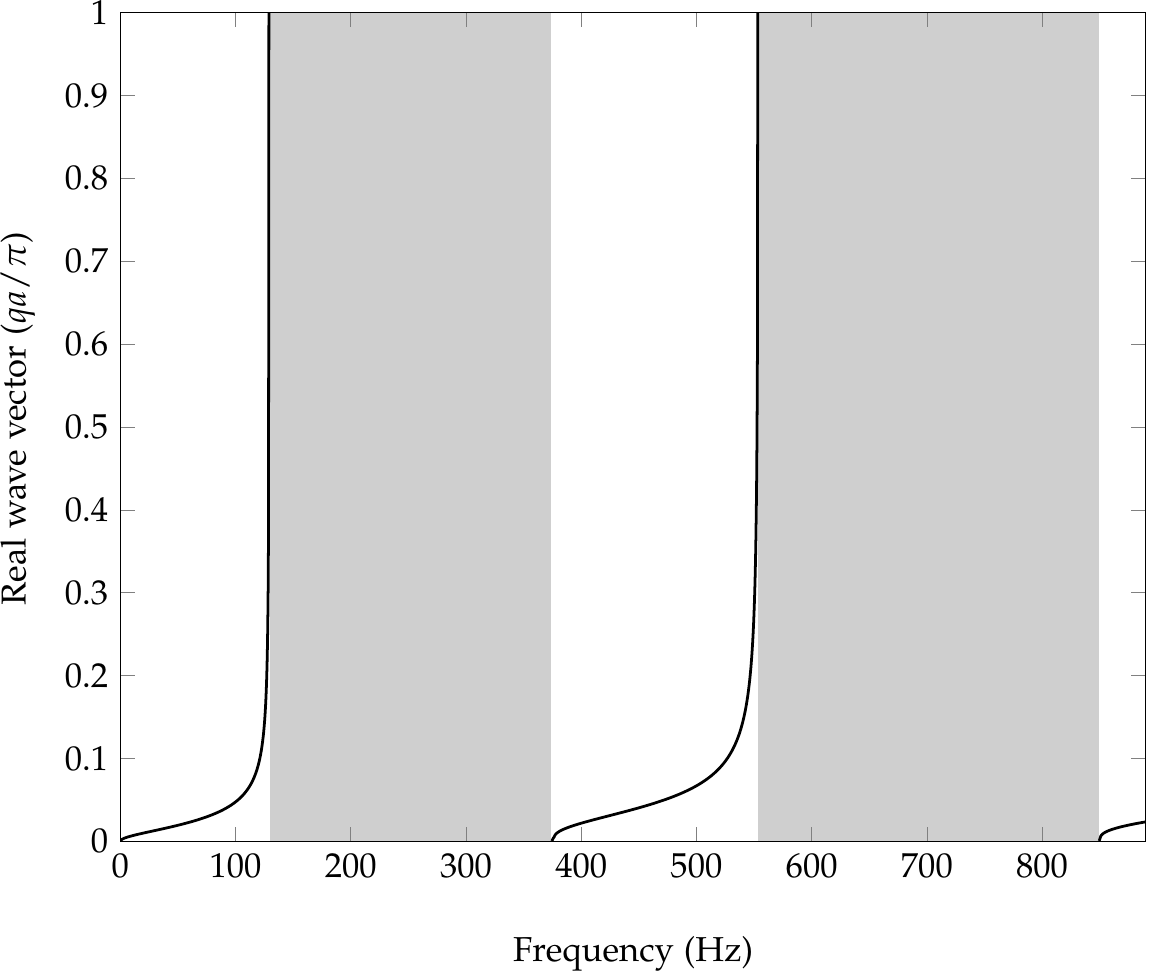}
\caption{\label{fig:qvec2}Band gaps of the infinite locally-resonant sandwich beam in Figure~\ref{fig:figure2} with 2-DOF resonators.}
\end{figure}

Now, consider the locally-resonant sandwich beam with 2-DOF resonators. The band gaps of the infinite beam without damping, calculated by the transfer matrix approach \cite{liu2007design}, are reported in Figure~\ref{fig:qvec2}. As expected, there are two band gaps, over the frequency ranges 130-\SI{374}{\Hz} and 553-\SI{849}{\Hz}. 
For the finite beam with damping, the first 161 complex eigenvalues calculated by the contour-integral algorithm in Section 4.1 are reported in Table~\ref{tab:eigen_2dof}. 
Again, the algorithm proves capable of capturing several eigenvalues close to each other, some differing even by a few digits, as a result of local resonance. 
\begin{figure}[h]
\centering
\includegraphics[scale=0.545]{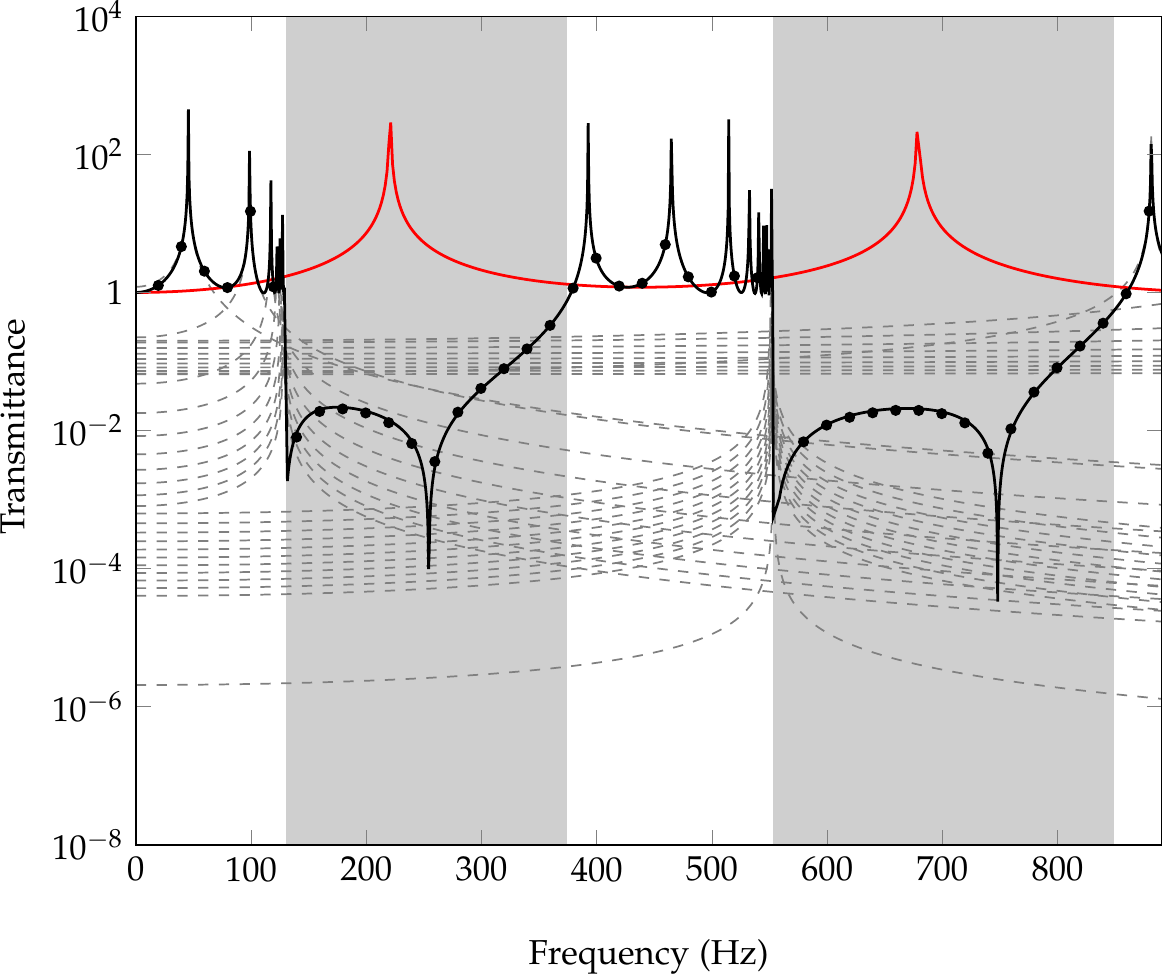}
\caption{\label{fig:trs_2dof}Transmittance of the cantilever locally-resonant sandwich beam in Figure 2 with 2-DOF resonators: exact response \eqref{eqn:exact_FRF} (black continuous line); total modal response \eqref{eqn:modal_FRF} for M = 161 (black dots); single modal responses \eqref{eqn:modal_FRF_contr} (gray dashed lines) for $k = 1,...,10$, $k=40,...,50$ and $k= 60,...,70$; exact response without resonators (red continuous line).}
\end{figure}

\begin{figure}[h]
\centering
\captionsetup[subfigure]{labelformat=empty,justification=centering}
\begin{subfigure}{\linewidth}
\centering
\includegraphics[scale=0.537]{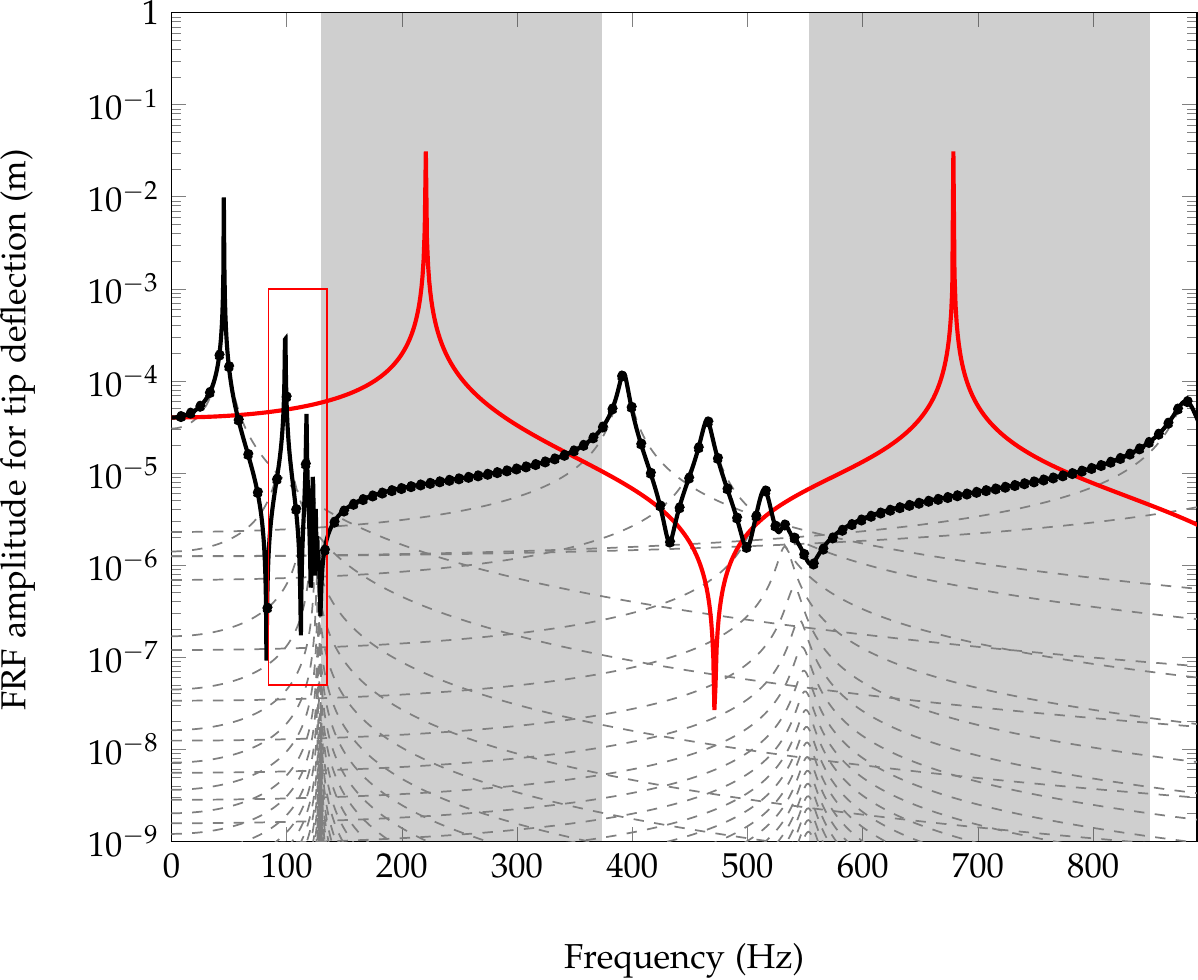}
\caption{\label{fig:frf_2dof_a}~~~~~~~~~~~~~~~~(a)}
\end{subfigure}
\begin{subfigure}{\linewidth}
\centering
\includegraphics[scale=0.537]{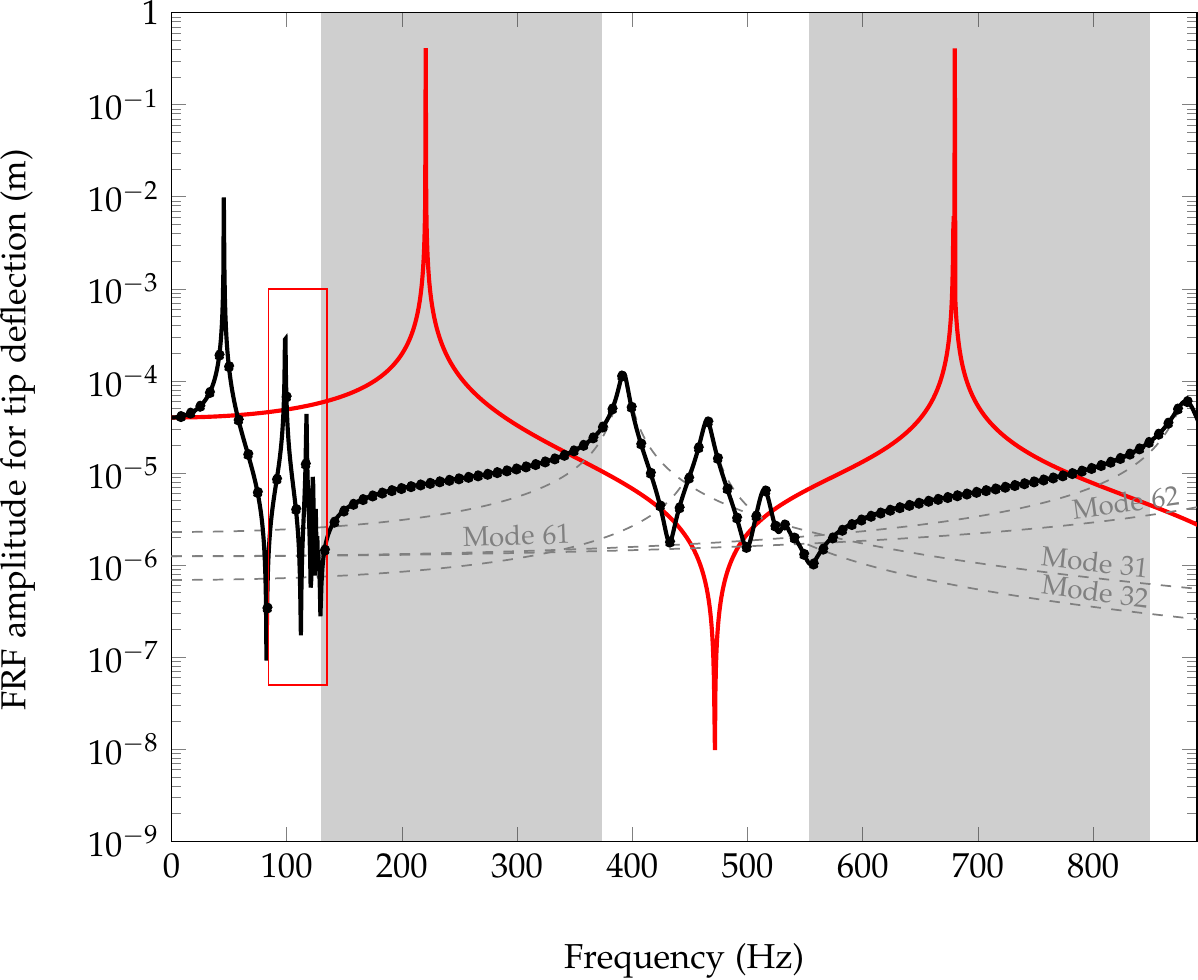}
\caption{\label{fig:frf_2dof_b}~~~~~~~~~~~~~~~~(b)}
\end{subfigure}
\begin{subfigure}{\linewidth}
\centering
\includegraphics[scale=0.537]{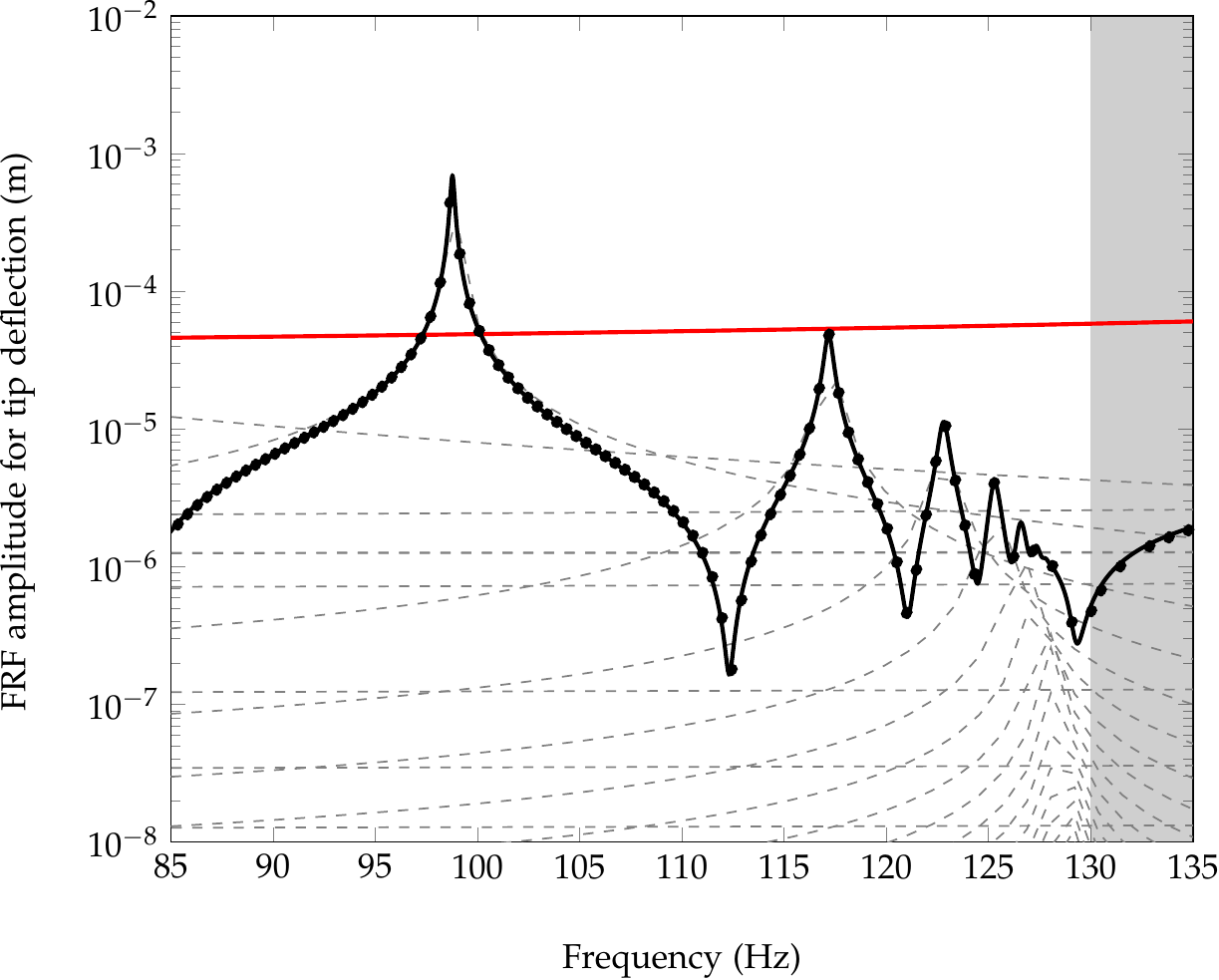}
\caption{\label{fig:frf_2dof_c}~~~~~~~~~~~~~~~~(c)}
\end{subfigure}
\caption{\label{fig:frf_2dof}FRF for tip deflection of cantilever locally-resonant sandwich beam in Figure~\ref{fig:figure2} with 2-DOF resonators, under a unit harmonic force applied at the free end: exact response \eqref{eqn:exact_FRF} (black continuous line); total modal response \eqref{eqn:modal_FRF} for $M = 161$ (black dots); single modal responses \eqref{eqn:modal_FRF_contr} (gray dashed lines); exact response without resonators (red continuous line); modal responses \eqref{eqn:modal_FRF_contr} are reported for $k = 1,...,52$ and $k=60,...,62$ (Fig.~\ref{fig:frf_2dof_a}) and $k=31, 32, 61, 62$ (Fig.~\ref{fig:frf_2dof_b}); a zoomed view is included (Fig.~\ref{fig:frf_2dof_c}).}
\end{figure}

For a further insight, transmittance and FRF for the tip deflection under a unit harmonic force at the free end are reported in Figure~\ref{fig:trs_2dof} and Figure~\ref{fig:frf_2dof}, respectively. Again, the exact solution \eqref{eqn:exact_FRF} and the modal expansion \eqref{eqn:modal_FRF} are in perfect agreement, proving the correctness of the two approaches. In this case, the modal expansion \eqref{eqn:modal_FRF} represents very accurately both the transmittance and the FRF with $M=161$ over the frequency domain 0-\SI{890}{\Hz} (see Figure~\ref{fig:frf_2dof_a} and zoomed view in Figure~\ref{fig:frf_2dof_c}). Further comments mirror those made for the beam with 1-DOF resonators, i.e.: the transmittance within the band gaps is a few orders of magnitude lower than the transmittance over the remaining frequency domain, meaning that the wave attenuation properties of the infinite beam hold also for the finite beam; there are no resonance modes within the two band gaps. 

\begin{figure}[h]
\captionsetup[subfigure]{labelformat=empty,justification=centering}
\begin{subfigure}{\linewidth}
\centering
\includegraphics[scale=0.545]{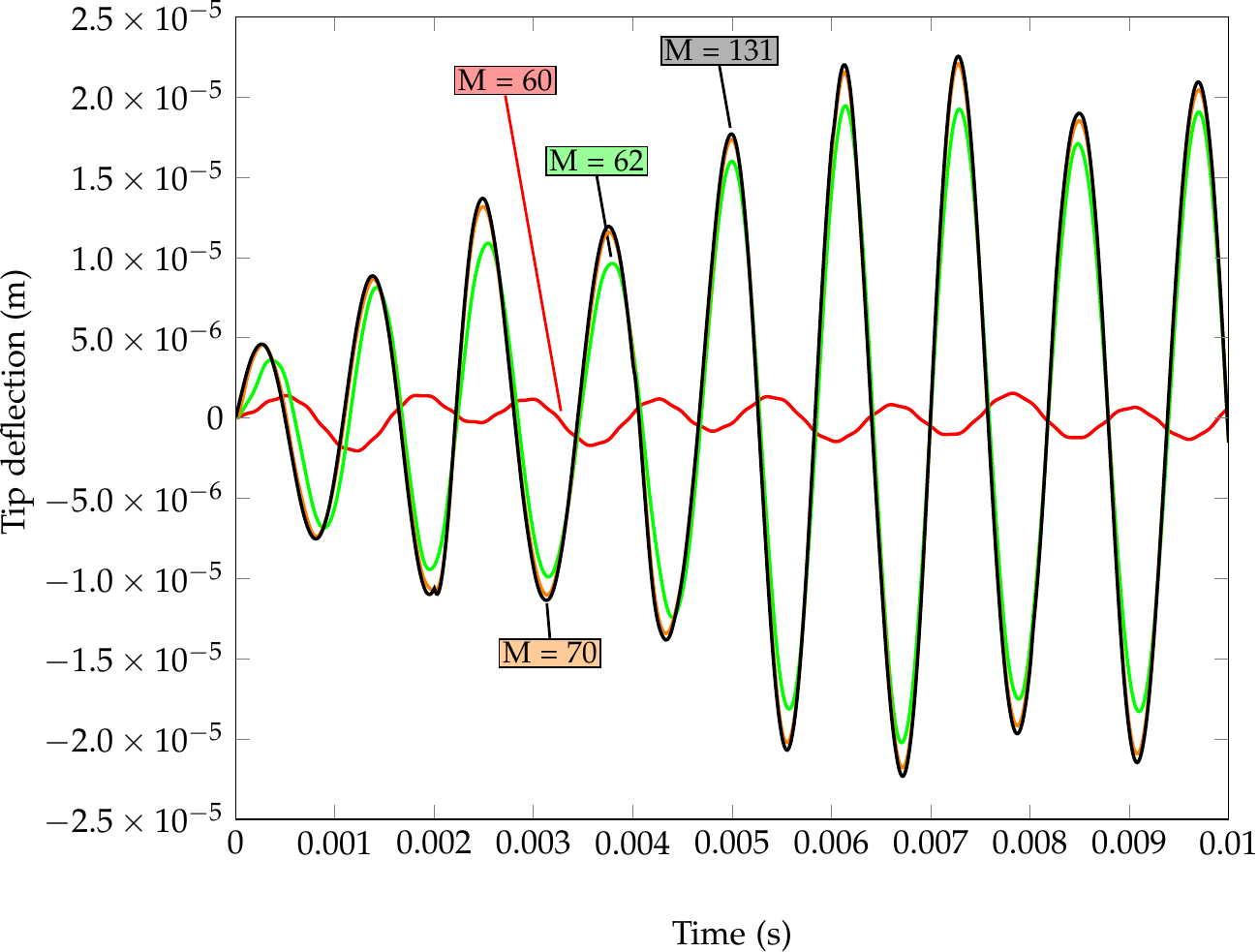}
\caption{\label{fig:irf_2dof_a}~~~~~~~~~~~~~~~~(a)}
\end{subfigure}
\begin{subfigure}{\linewidth}
\centering
\includegraphics[scale=0.545]{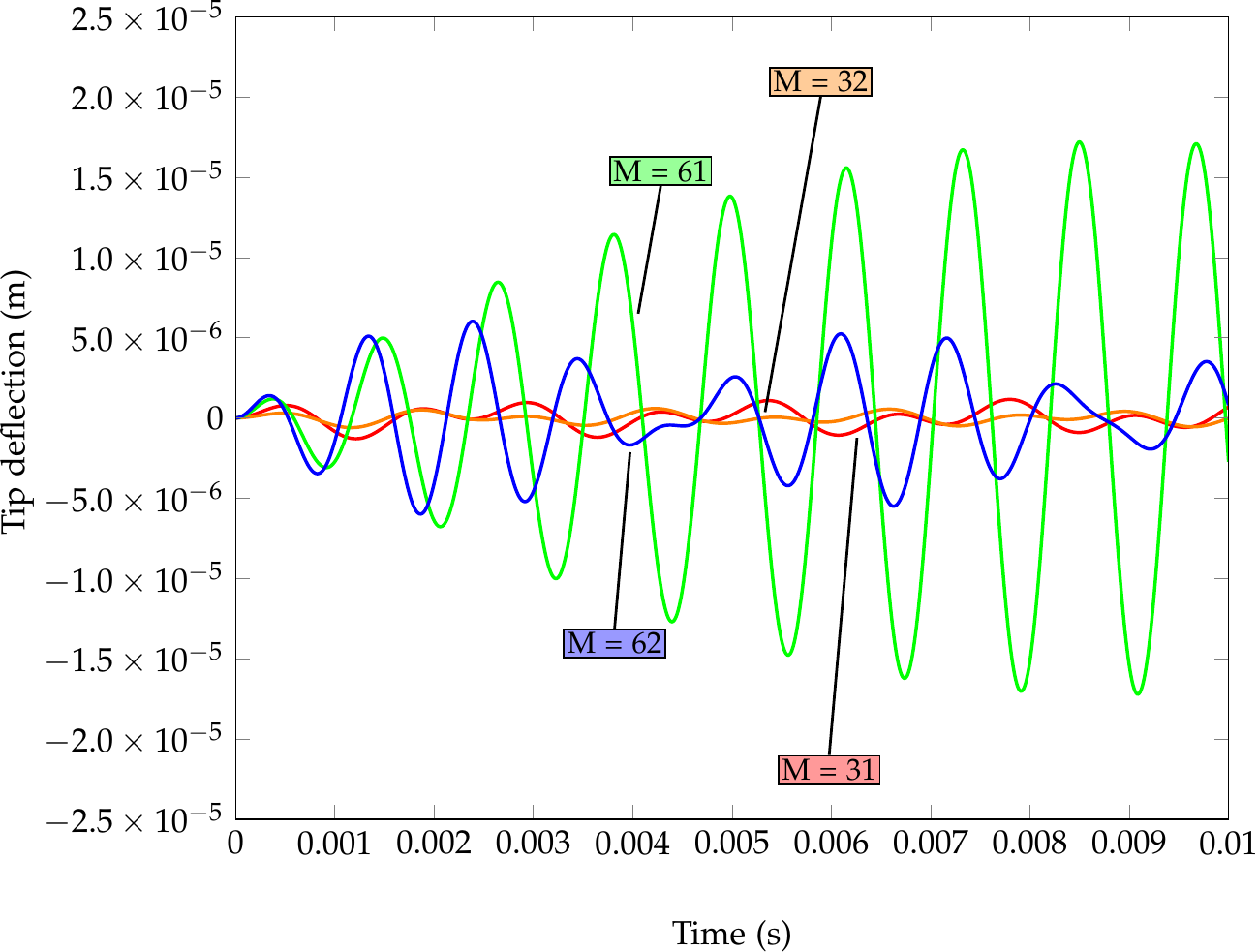}
\caption{\label{fig:irf_2dof_b}~~~~~~~~~~~~~~~~(b)}
\end{subfigure}
\captionsetup[subfigure]{labelformat=empty,justification=centering}
\begin{subfigure}{\linewidth}
\centering
\includegraphics[scale=0.545]{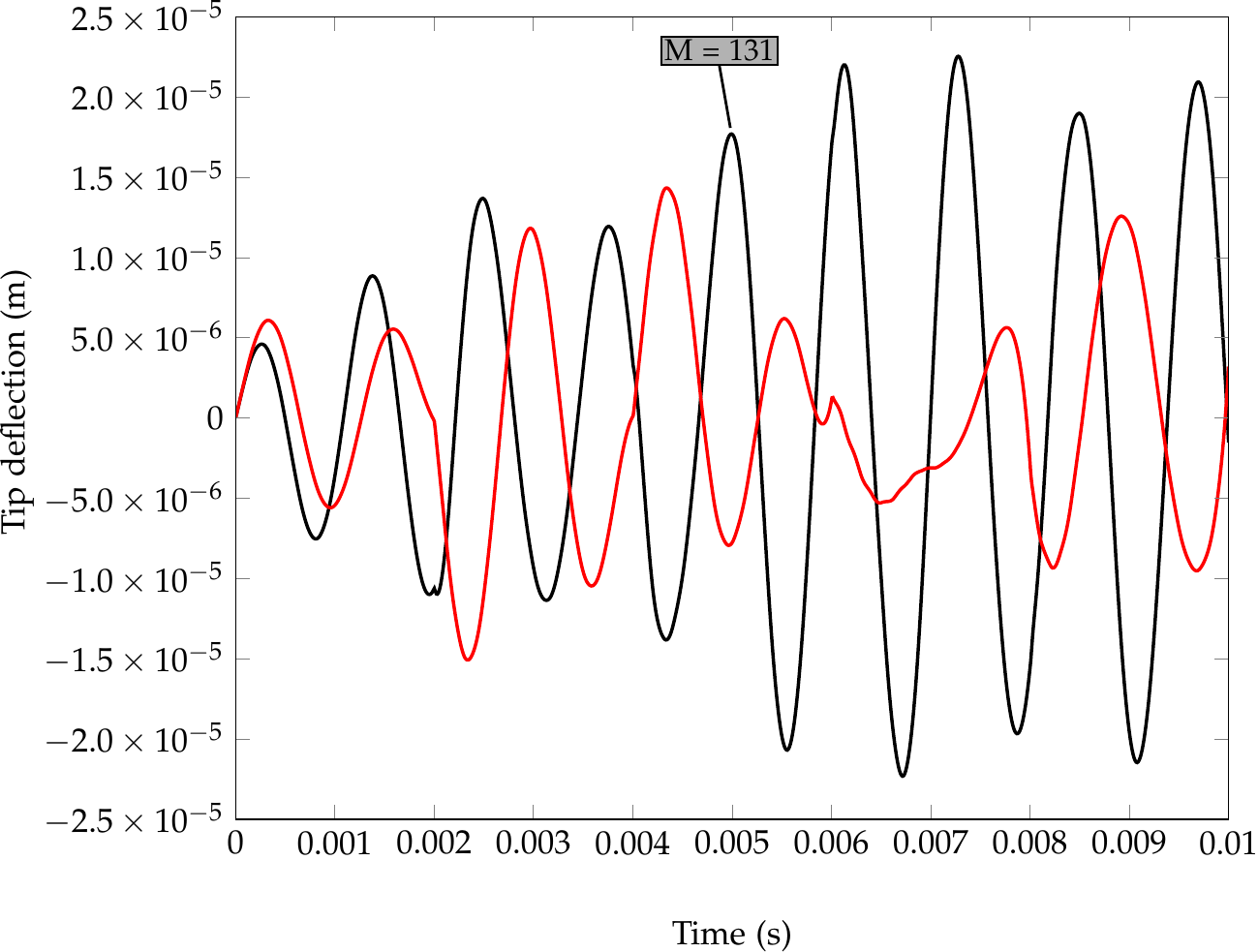}
\caption{\label{fig:irf_2dof_c}~~~~~~~~~~~~~~~~(c)}
\end{subfigure}
\caption{\label{fig:irf_2dof} Tip deflection of the cantilever locally-resonant sandwich beam in Figure~\ref{fig:figure2} for 2-DOF resonators, under a unit cosine force with frequency 830 Hz: (a) total response for increasing number of modes $M$ in Eq.~\eqref{eqn:modal_IRF}; (b) single modal response for most significant modes; (c) total response for $M=161$ in Eq.~\eqref{eqn:modal_IRF} (black continuous line) and response of the beam without resonators (red continuous line).}
\end{figure}

The FRF of the beam with resonators is generally lower than the corresponding one without resonators within the two band gaps, except for a limited frequency range at the vicinity of the right end of the second band gap. Figure \ref{fig:frf_2dof_b} shows that the most significant contributions to the FRF at the right end of the second bandgap are associated with modes 31-32-61-62. This result is confirmed by the time analysis of the tip deflection under a unit cosine force applied at the free end, with frequency 830~Hz, reported in Figure~\ref{fig:irf_2dof}. Indeed, the response built using Eq.~\eqref{eqn:modal_IRF} for the IRF attains the same order of magnitude of the response of the beam without resonators due mainly to the contributions of these modes; on the other hand, no significant changes in the time response are noticed if more than $M=70$ modes are included.

The final step is to compare the FRF in Figure~\ref{fig:frf_2dof} with the corresponding one obtained when the resonators are considered as exerting distributed forces over the mutual distance $a$, reported in Figure~\ref{fig:frf_2dof_distr} (for the calculation see Appendix~B). As for the locally-resonant sandwich beam with 1-DOF resonators, no significant differences are encountered between the FRFs obtained by the two models.

\begin{figure}[h]
\centering
\captionsetup[subfigure]{labelformat=empty,justification=centering}
\begin{subfigure}{\columnwidth}
\centering
\includegraphics[scale=0.545]{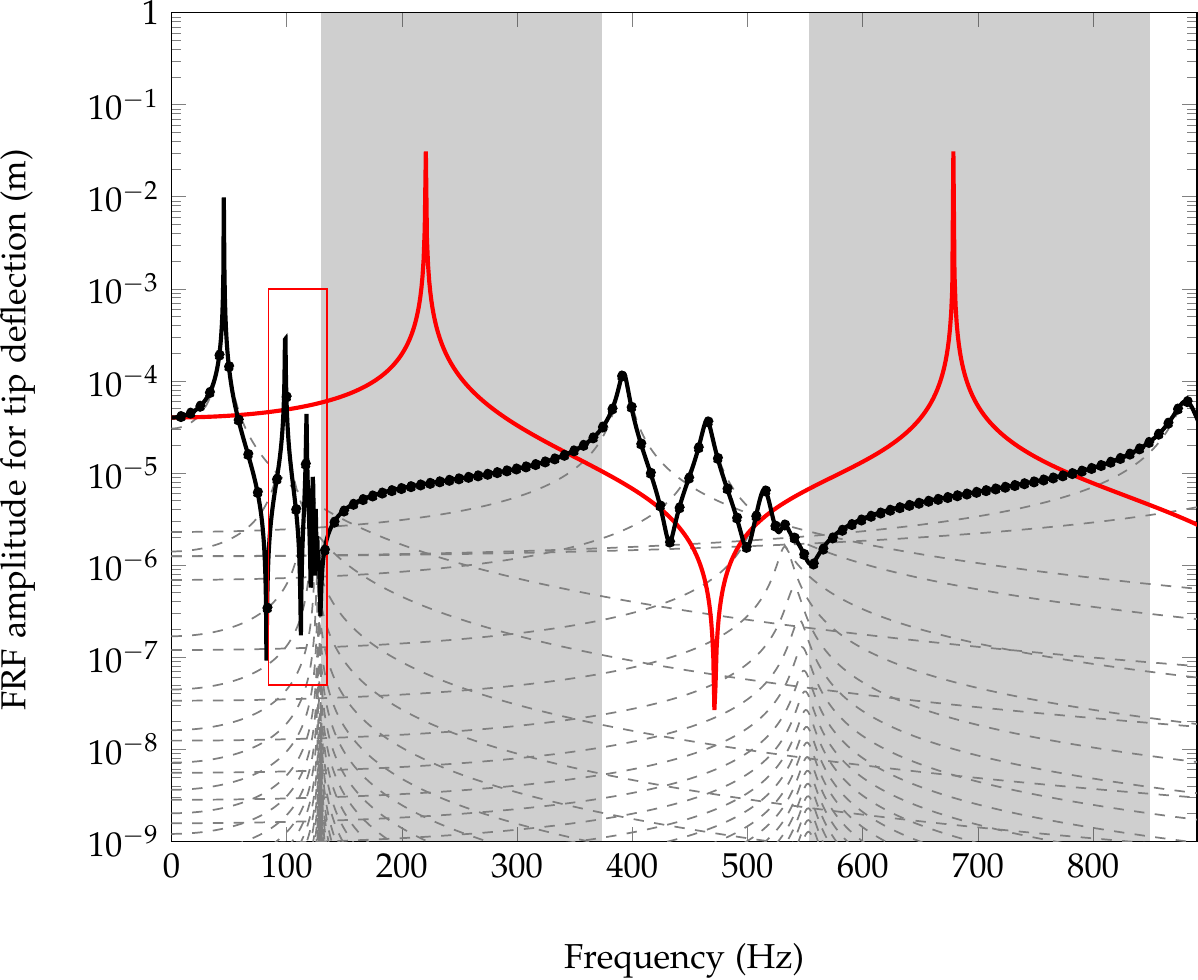}
\caption{\label{fig:frf_2dof_distr_a}~~~~~~~~~~~~~~~~(a)}
\end{subfigure}
\begin{subfigure}{\columnwidth}
\centering
\includegraphics[scale=0.545]{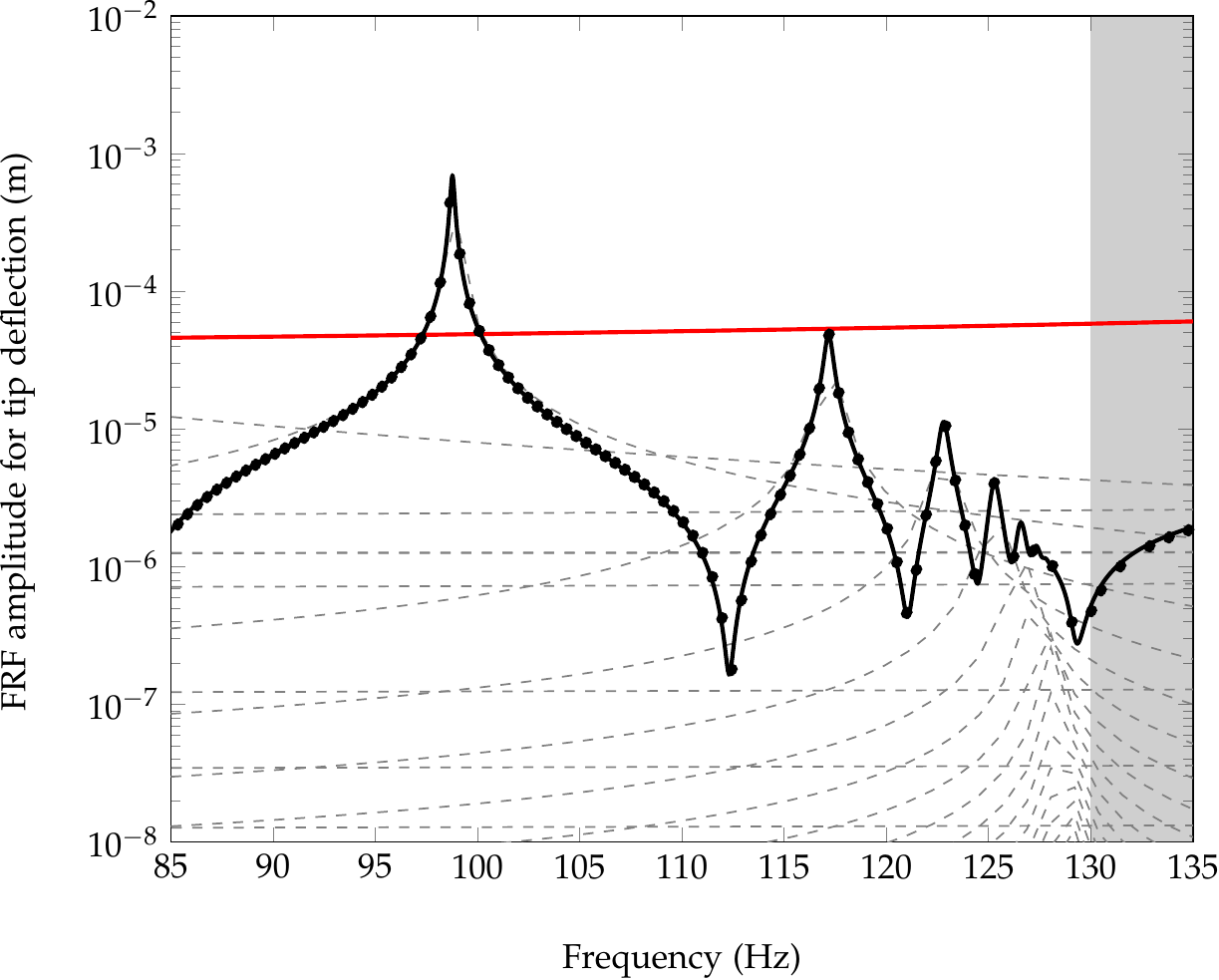}
\caption{\label{fig:frf_2dof_distr_b}~~~~~~~~~~~~~~~~(b)}
\end{subfigure}
\caption{\label{fig:frf_2dof_distr} FRF for tip deflection of cantilever locally-resonant sandwich beam in Figure~\ref{fig:figure2} with 2-DOF resonators modelled as exerting distributed forces over the mutual distance $a$, under a unit harmonic force applied at the free end: exact response \eqref{eqn:exact_FRF} (black continuous line); total modal response \eqref{eqn:modal_FRF} for $M = 161$ (black dots); single modal responses \eqref{eqn:modal_FRF_contr} (gray dashed lines); exact response without resonators (red continuous line); modal responses \eqref{eqn:modal_FRF_contr} are reported for $k=1,...,52$ and $k=60,...,62$ (Fig.~\ref{fig:frf_2dof_distr_a}); a zoomed view is included (Fig.~\ref{fig:frf_2dof_distr_b}).}
\end{figure}
\clearpage
\begin{table*}
\resizebox{\textwidth}{!}{
\begin{tabular}{cl}
\toprule
Mode&Eigenvalue\\
\midrule
$1$ &$  286.280\pm0.027\mathrm{i}$\\ 
$2$ &$  620.619\pm0.632\mathrm{i}$\\ 
$3$ &$  736.251\pm1.280\mathrm{i}$\\ 
$4$ &$  771.551\pm1.552\mathrm{i}$\\ 
$5$ &$  787.217\pm1.686\mathrm{i}$\\ 
$6$ &$  795.224\pm1.757\mathrm{i}$\\ 
$7$ &$  799.919\pm1.800\mathrm{i}$\\ 
$8$ &$  802.868\pm1.827\mathrm{i}$\\ 
$9$ &$  804.854\pm1.846\mathrm{i}$\\ 
$10$ &$  806.244\pm1.859\mathrm{i}$\\ 
$11$ &$  807.260\pm1.868\mathrm{i}$\\ 
$12$ &$  808.020\pm1.876\mathrm{i}$\\ 
$13$ &$  808.604\pm1.881\mathrm{i}$\\ 
$14$ &$  809.062\pm1.886\mathrm{i}$\\ 
$15$ &$  809.426\pm1.889\mathrm{i}$\\ 
$16$ &$  809.720\pm1.892\mathrm{i}$\\ 
$17$ &$  809.960\pm1.894\mathrm{i}$\\ 
$18$ &$  810.158\pm1.896\mathrm{i}$\\ 
$19$ &$  810.322\pm1.898\mathrm{i}$\\ 
$20$ &$  810.458\pm1.899\mathrm{i}$\\ 
$21$ &$  810.572\pm1.900\mathrm{i}$\\ 
$22$ &$  810.668\pm1.901\mathrm{i}$\\ 
$23$ &$  810.747\pm1.902\mathrm{i}$\\ 
$24$ &$  810.814\pm1.902\mathrm{i}$\\ 
$25$ &$  810.868\pm1.903\mathrm{i}$\\ 
$26$ &$  810.912\pm1.903\mathrm{i}$\\ 
$27$ &$  810.947\pm1.904\mathrm{i}$\\ 
$28$ &$  810.973\pm1.904\mathrm{i}$\\ 
$29$ &$  810.992\pm1.904\mathrm{i}$\\ 
$30$ &$  811.003\pm1.904\mathrm{i}$\\ 
$31$ &$  2466.989\pm23.532\mathrm{i}$\\ 
$32$ &$  2920.847\pm27.657\mathrm{i}$\\ 
$33$ &$  3233.359\pm33.001\mathrm{i}$\\ 
$34$ &$  3345.749\pm35.461\mathrm{i}$\\ 
$35$ &$  3397.235\pm36.666\mathrm{i}$\\ 
$36$ &$  3423.788\pm37.305\mathrm{i}$\\ 
$37$ &$  3439.413\pm37.685\mathrm{i}$\\ 
$38$ &$  3449.241\pm37.926\mathrm{i}$\\ 
$39$ &$  3455.865\pm38.089\mathrm{i}$\\ 
$40$ &$  3460.505\pm38.204\mathrm{i}$\\ 
$41$ &$  3463.895\pm38.287\mathrm{i}$\\ 
\bottomrule
\end{tabular}
\begin{tabular}{cl}
\toprule
Mode&Eigenvalue\\
\midrule
$42$ &$  3466.432\pm38.350\mathrm{i}$\\ 
$43$ &$  3468.384\pm38.399\mathrm{i}$\\ 
$44$ &$  3469.911\pm38.437\mathrm{i}$\\ 
$45$ &$  3471.128\pm38.467\mathrm{i}$\\ 
$46$ &$  3472.110\pm38.491\mathrm{i}$\\ 
$47$ &$  3472.912\pm38.511\mathrm{i}$\\ 
$48$ &$  3473.571\pm38.528\mathrm{i}$\\ 
$49$ &$  3474.118\pm38.541\mathrm{i}$\\ 
$50$ &$  3474.574\pm38.553\mathrm{i}$\\ 
$51$ &$  3474.956\pm38.562\mathrm{i}$\\ 
$52$ &$  3475.274\pm38.570\mathrm{i}$\\ 
$53$ &$  3475.541\pm38.577\mathrm{i}$\\ 
$54$ &$  3475.762\pm38.582\mathrm{i}$\\ 
$55$ &$  3475.944\pm38.587\mathrm{i}$\\ 
$56$ &$  3476.393\pm38.598\mathrm{i}$\\ 
$57$ &$  3476.357\pm38.597\mathrm{i}$\\ 
$58$ &$  3476.295\pm38.595\mathrm{i}$\\ 
$59$ &$  3476.207\pm38.593\mathrm{i}$\\ 
$60$ &$  3476.091\pm38.590\mathrm{i}$\\ 
$61$ &$  5537.644\pm58.228\mathrm{i}$\\ 
$62$ &$  6647.837\pm53.665\mathrm{i}$\\ 
$63$ &$  9073.851\pm47.573\mathrm{i}$\\ 
$64$ &$  11819.176\pm44.797\mathrm{i}$\\ 
$65$ &$  14783.101\pm43.456\mathrm{i}$\\ 
$66$ &$  17779.491\pm42.718\mathrm{i}$\\ 
$67$ &$  20842.056\pm42.297\mathrm{i}$\\ 
$68$ &$  23904.008\pm41.994\mathrm{i}$\\ 
$69$ &$  26998.851\pm41.805\mathrm{i}$\\ 
$70$ &$  30083.321\pm41.607\mathrm{i}$\\ 
$71$ &$  33188.621\pm41.451\mathrm{i}$\\ 
$72$ &$  36266.575\pm41.032\mathrm{i}$\\ 
$73$ &$  39326.373\pm39.566\mathrm{i}$\\ 
$74$ &$  41661.862\pm15.667\mathrm{i}$\\ 
$75$ &$  42965.366\pm30.348\mathrm{i}$\\ 
$76$ &$  45787.637\pm40.462\mathrm{i}$\\ 
$77$ &$  48876.968\pm41.055\mathrm{i}$\\ 
$78$ &$  51980.110\pm41.218\mathrm{i}$\\ 
$79$ &$  55101.302\pm41.241\mathrm{i}$\\ 
$80$ &$  58220.947\pm41.251\mathrm{i}$\\ 
$81$ &$  61347.886\pm41.238\mathrm{i}$\\ 
$82$ &$  64472.968\pm41.223\mathrm{i}$\\ 
\bottomrule
\end{tabular}
\begin{tabular}{cl}
\toprule
Mode&Eigenvalue\\
\midrule 
$83$ &$  67602.459\pm41.199\mathrm{i}$\\ 
$84$ &$  70730.285\pm41.172\mathrm{i}$\\ 
$85$ &$  73861.305\pm41.136\mathrm{i}$\\ 
$86$ &$  76990.732\pm41.089\mathrm{i}$\\ 
$87$ &$  80122.759\pm41.024\mathrm{i}$\\ 
$88$ &$  83253.055\pm40.924\mathrm{i}$\\ 
$89$ &$  86385.636\pm40.751\mathrm{i}$\\ 
$90$ &$  89515.805\pm40.352\mathrm{i}$\\ 
$91$ &$  92646.501\pm38.420\mathrm{i}$\\ 
$92$ &$  95781.756\pm44.114\mathrm{i}$\\ 
$93$ &$  98908.259\pm42.056\mathrm{i}$\\ 
$94$ &$  101966.090\pm37.151\mathrm{i}$\\ 
$95$ &$  102806.610\pm5.535\mathrm{i}$\\ 
$96$ &$  105213.896\pm41.472\mathrm{i}$\\ 
$97$ &$  108336.858\pm41.610\mathrm{i}$\\ 
$98$ &$  111465.571\pm41.591\mathrm{i}$\\ 
$99$ &$  114599.106\pm41.550\mathrm{i}$\\ 
$100$ &$  117731.872\pm41.522\mathrm{i}$\\ 
$101$ &$  120866.478\pm41.495\mathrm{i}$\\ 
$102$ &$  124000.350\pm41.475\mathrm{i}$\\ 
$103$ &$  127135.375\pm41.456\mathrm{i}$\\ 
$104$ &$  130269.785\pm41.442\mathrm{i}$\\ 
$105$ &$  133405.068\pm41.427\mathrm{i}$\\ 
$106$ &$  136539.800\pm41.415\mathrm{i}$\\ 
$107$ &$  139675.274\pm41.402\mathrm{i}$\\ 
$108$ &$  142810.201\pm41.392\mathrm{i}$\\ 
$109$ &$  145945.818\pm41.380\mathrm{i}$\\ 
$110$ &$  149080.823\pm41.369\mathrm{i}$\\ 
$111$ &$  152216.521\pm41.357\mathrm{i}$\\ 
$112$ &$  155351.390\pm41.344\mathrm{i}$\\ 
$113$ &$  158486.963\pm41.328\mathrm{i}$\\ 
$114$ &$  161620.773\pm41.300\mathrm{i}$\\ 
$115$ &$  164753.712\pm41.215\mathrm{i}$\\ 
$116$ &$  167637.727\pm11.615\mathrm{i}$\\ 
$117$ &$  167994.170\pm29.950\mathrm{i}$\\ 
$118$ &$  171039.168\pm41.172\mathrm{i}$\\ 
$119$ &$  174172.626\pm41.192\mathrm{i}$\\ 
$120$ &$  177306.906\pm41.164\mathrm{i}$\\ 
$121$ &$  180442.694\pm41.071\mathrm{i}$\\ 
$122$ &$  183578.055\pm40.860\mathrm{i}$\\ 
$123$ &$  186713.794\pm39.818\mathrm{i}$\\ 
\bottomrule
\end{tabular}
\begin{tabular}{cl}
\toprule
Mode&Eigenvalue\\
\midrule 
$124$ &$  189850.729\pm42.944\mathrm{i}$\\ 
$125$ &$  192986.582\pm41.899\mathrm{i}$\\ 
$126$ &$  196122.458\pm41.691\mathrm{i}$\\ 
$127$ &$  199258.737\pm41.600\mathrm{i}$\\ 
$128$ &$  202394.761\pm41.545\mathrm{i}$\\ 
$129$ &$  205531.118\pm41.513\mathrm{i}$\\ 
$130$ &$  208667.207\pm41.488\mathrm{i}$\\ 
$131$ &$  211803.611\pm41.470\mathrm{i}$\\ 
$132$ &$  214939.706\pm41.454\mathrm{i}$\\ 
$133$ &$  218076.128\pm41.442\mathrm{i}$\\ 
$134$ &$  221212.123\pm41.430\mathrm{i}$\\ 
$135$ &$  224348.457\pm41.419\mathrm{i}$\\ 
$136$ &$  227483.857\pm41.401\mathrm{i}$\\ 
$137$ &$  230618.697\pm41.353\mathrm{i}$\\ 
$138$ &$  233427.574\pm4.545\mathrm{i}$\\ 
$139$ &$  233798.678\pm37.005\mathrm{i}$\\ 
$140$ &$  236898.785\pm41.357\mathrm{i}$\\ 
$141$ &$  240034.101\pm41.377\mathrm{i}$\\ 
$142$ &$  243169.755\pm41.377\mathrm{i}$\\ 
$143$ &$  246306.196\pm41.372\mathrm{i}$\\ 
$144$ &$  249442.418\pm41.366\mathrm{i}$\\ 
$145$ &$  252578.998\pm41.358\mathrm{i}$\\ 
$146$ &$  255715.386\pm41.350\mathrm{i}$\\ 
$147$ &$  258852.019\pm41.339\mathrm{i}$\\ 
$148$ &$  261988.485\pm41.327\mathrm{i}$\\ 
$149$ &$  265125.149\pm41.310\mathrm{i}$\\ 
$150$ &$  268261.659\pm41.287\mathrm{i}$\\ 
$151$ &$  271398.349\pm41.253\mathrm{i}$\\ 
$152$ &$  274534.881\pm41.192\mathrm{i}$\\ 
$153$ &$  277671.594\pm41.054\mathrm{i}$\\ 
$154$ &$  280808.160\pm40.369\mathrm{i}$\\ 
$155$ &$  283944.735\pm42.413\mathrm{i}$\\ 
$156$ &$  287081.233\pm41.728\mathrm{i}$\\ 
$157$ &$  290217.891\pm41.589\mathrm{i}$\\ 
$158$ &$  293353.991\pm41.522\mathrm{i}$\\ 
$159$ &$  296489.787\pm41.462\mathrm{i}$\\ 
$160$ &$  299411.374\pm5.958\mathrm{i}$\\ 
$161$ &$  299665.134\pm35.609\mathrm{i}$\\
\\
\\
\\
\bottomrule
\end{tabular}
}
\caption{\label{tab:eigen_2dof}Complex eigenvalues of the cantilever locally-resonant sandwich beam in Figure~\ref{fig:figure2} with 2-DOF resonators}
\end{table*}

\clearpage
\section{Concluding Remarks}

The subject of this paper is the dynamics of locally-res\-o\-nant sandwich beams, featuring a periodic distribution of multi-DOF viscously-damped resonators within the core matrix. Modelling the system as an equivalent single-layer Timoshenko beam coupled with mass-spring-dashpot subsystems representing the resonators, exact closed analytical forms have been obtained for the frequency response, the modal impulse and frequency response functions. The solutions are built considering the resonators as exerting point forces and using the theory of generalized functions to handle the associated shear-force discontinuities; simple modifications, however, are required to include the alternative model of resonators exerting distributed forces over the mutual distance. The proposed modal analysis approach relies on pertinent orthogonality conditions for the complex modes and a recently-introduced contour-integral algorithm to tackle the challenging issues of calculating all complex eigenvalues, without missing anyone. Specifically, the eigenvalues are obtained from a characteristic equation built as determinant of an exact frequency-response matrix, whose size is $4\times 4$ regardless of the number of resonators. Numerical applications show exactness and robustness of the proposed solutions, showing their suitability for practical purposes.

\section{Acknowledgements}

The authors acknowledge financial support from the Italian Ministry of Education, University and Research (MIUR) under the `Departments of Excellence' grant L.232/2016. FF acknowledges financial support from MIUR under the PRIN 2017 National Grant `Multiscale Innovative Materials and Structures' (grant number 2017J4EAYB). 

\section{Appendix A}\label{app:A}

The matrix $\vect{\Omega}$ associated with the homogeneous solution of Eq.~\eqref{eqn:single_motion} is given as
\begin{equation}\label{eqn:om_mat}
\resizebox{\columnwidth}{!}{
$
\vect{\Omega}(x) = \begin{bmatrix}
\alpha_1 \mathrm{e}^{\lambda_1 x}&\alpha_2 \mathrm{e}^{\lambda_2 x}&\alpha_3 \mathrm{e}^{\lambda_3 x}&\alpha_4 \mathrm{e}^{\lambda_4 x}\\
\mathrm{e}^{\lambda_1 x}&\mathrm{e}^{\lambda_2 x}&\mathrm{e}^{\lambda_3 x}&\mathrm{e}^{\lambda_4 x}\\
 \mathrm{e}^{\lambda_1 x}\kappa GA(1+\alpha_1)& \mathrm{e}^{\lambda_2 x}\kappa GA(1+\alpha_2)& \mathrm{e}^{\lambda_3 x}\kappa GA(1+\alpha_3)& \mathrm{e}^{\lambda_4 x}\kappa GA(1+\alpha_4)\\
EI\mathrm{e}^{\lambda_1 x}&EI\mathrm{e}^{\lambda_2 x}&EI\mathrm{e}^{\lambda_3 x}&EI \mathrm{e}^{\lambda_4 x}\\
\end{bmatrix}
$}
\end{equation}
where $\alpha_i$ ($i=1,...,4$) is given by Eq.~\eqref{eqn:alpha}.

The FRF vector $\vect{Y}(x,\omega)$ can be written as
\begin{equation}\label{eqn:exact_FRF_resonators}
\vect{Y}(x,\omega) = \vect{\Omega}(x,\omega)\vect{c} + \vect{R}(x)\vect{\Lambda}(\omega) + \widetilde{\vect{Y}}^{(f)}(x)
\end{equation}
where vector $\widetilde{\vect{Y}}^{(f)}$ and matrix $\vect{R}$ are given by
\begin{equation}
\begin{aligned}
&\widetilde{\vect{Y}}^{(f)}(x) = \int_{0}^{L} \vect{J}(x,y)f_{v}(y)~\mathrm{d}y\\ 
&\vect{R}(x)= \begin{bmatrix}
\vect{J}(x,x_1)&\dots&\vect{J}(x,x_N)
\end{bmatrix}
\end{aligned}
\end{equation}
being $\vect{J}(x,x_j)$ defined as
\begin{equation}\label{eqn:J_vector}
\vect{J}(x,x_j) = \begin{bmatrix}
J_{V}(x,x_j)\\
J_{\Phi}(x,x_j)\\
J_{T}(x,x_j)\\
J_{M}(x,x_j)\\
\end{bmatrix}
\end{equation}
with:
\begin{equation}\label{eqn:J_M_T}
J_{T}(x,x_j) = GA \left(\gtder{J_{V}}{x} + J_{\Phi} \right); \quad J_{M}(x,x_j) = EI \gtder{J_{\Phi}}{x}
\end{equation}
In Eq.~\eqref{eqn:exact_FRF_resonators}, $\vect{\Lambda}$ is a vector collecting the unknown reaction forces $R_j$ of the resonators and satisfying the following linear system
\begin{equation}\label{eqn:reac_sys}
\vect{\Lambda} = \vect{\Phi}_{\Omega}\vect{c} + \vect{\Phi}_{J}\vect{\Lambda} + \vect{\Phi}_{f} 
\end{equation}
where $\vect{\Phi}_{\Omega}$ is matrix whose $j^{\mathrm{th}}$ row is the first row of the matrix $\vect{\Omega}$ evaluated at $x_j$, i.e. $\vect{\Omega}_{1}(x_j)$, hence 
\begin{equation}
\vect{\Phi}_{\Omega} =
-k_{eq}(\omega)
\begin{bmatrix}
\vect{\Omega}_{1}(x_1)\\
\vdots\\
\vect{\Omega}_{1}(x_N)
\end{bmatrix}
\end{equation}
In Eq.~\eqref{eqn:reac_sys}, $\vect{\Phi}_{J}$ is the strict lower triangular matrix
\begin{equation}
\vect{\Phi}_{J}= -k_{eq}(\omega)\begin{bmatrix}
0&0&\dots&0\\
J_{V}(x_2,x_1)&0&\dots&0\\
\vdots&\ddots&&\vdots\\
J_{V}(x_N,x_1)&\dots&J_{V}(x_N,x_{N-1})&0
\end{bmatrix}
\end{equation}
and $\vect{\Phi}_f$ is a vector containing the first component of vector $\vect{Y}^{(f)}$ evaluated at $x_j$:
\begin{equation}
\vect{\Phi}_{f}= -k_{eq}(\omega)\begin{bmatrix}
Y^{(f)}_1(x_1)\\
\vdots\\
Y^{(f)}_1 (x_N)
\end{bmatrix}
\end{equation}
The solution of Eq.~\eqref{eqn:reac_sys} is given by
\begin{equation}
\vect{\Lambda} = (\vect{I} - \vect{\Phi}_{J})^{-1} (\vect{\Phi}_{\Omega}\vect{c} + \vect{\Phi}_f)
\end{equation}
where the inverse matrix $(\vect{I} - \vect{\Phi}_{J})^{-1}$ can be calculated in closed form as:
\begin{equation}
(\vect{I}-\vect{\Phi}_{J})^{-1} = \sum_{j=0}^{N-1}\vect{\Phi}^{\,j}_{J}
\end{equation}
Replacing Eq.~\eqref{eqn:reac_sys} for $\vect{\Lambda}$ in Eq.~\eqref{eqn:exact_FRF_resonators} leads to Eq.~\eqref{eqn:exact_FRF} of the main text, where matrix $\vect{W}$ is
\begin{equation}
\vect{W}(x,\omega) = \vect{\Omega}(x,\omega) + \vect{R}(x)(\vect{I} - \vect{\Phi}_{J})^{-1}\vect{\Phi}_{\Omega}
\end{equation}
and vector $\vect{Y}^{(f)}$ is
\begin{equation}
\vect{Y}^{(f)}(x) = \vect{R}(x)(\vect{I} - \vect{\Phi}_{J})^{-1}\vect{\Phi}_f + \widetilde{\vect{Y}}_{f}(x)  
\end{equation}
On the other hand, closed analytical expressions are available for vector $\vect{Y}^{(f)}$ in Eq.~\eqref{eqn:exact_FRF} using simple rules of integration of generalized functions \cite{failla2016exact}.

\section{Appendix B}\label{app:B}
Eq.~\eqref{eqn:exact_FRF}, Eq.~\eqref{eqn:modal_IRF} and Eq.~\eqref{eqn:modal_FRF} of the main text can be applied with little modifications also if the resonators are modelled as exerting distributed forces over the mutual distance $a$ \cite{chen2011containing,chen2011dynamic}. In this case, Eqs.~\eqref{eqn:p1}-\eqref{eqn:p2} become
\begin{small}
\begin{flalign}\label{eqn:p1_p2_distr}
&p_1 = (a EI GA)^{-1}[(a EI + a GI ) \rho A\omega^2 - EI  k_{eq}(\omega)]&&\\
&p_2 = (a EI GA)^{-1}[(\rho I \omega^2 - GA )a\rho A\omega^2 +(GA - \rho I \omega^2)k_{eq}(\omega)]&&
\end{flalign}
\end{small}
Eq.~\eqref{eqn:alpha} is
\begin{equation}\label{eqn:alpha_distr}
\alpha_i = \left\{
\begin{aligned}
&1& \quad&\text{if}\, Z = \Phi\\
&-\frac{a G A \lambda_i}{a (\rho A \omega^2 + GA \lambda_i^2) - k_{eq}(\omega)}&\quad&\text{if}\, Z = V
\end{aligned}
\right.
\end{equation}
The particular integrals Eqs.~\eqref{eqn:fundamental}-\eqref{eqn:fundamental_rotation} become
\begin{flalign*}
&J_{V}(x,x_j) = \tau \left[(\Theta _2 \Upsilon_1 + EI k_{eq})\sinh \left(2^{-1/2}\Theta _1 \left(x-x_j\right)\right) \right. &&\\  
& \left.  (\Theta _1 \Upsilon_2 - EI k_{eq})\sinh \left(2^{-1/2}\Theta _2 \left(x-x_j\right)\right) \right]\mathcal{H} \left(x-x_j\right)\\
&J_{\Phi}(x,x_j)= aGA \Xi_{1}^{-1}\left[\cosh\left(2^{-1/2}\Theta _2 \left(x-x_j\right)\right) \right. &&\\
&\left. - \cosh\left(2^{-1/2}\Theta _1 \left(x-x_j\right)\right)\right]\mathcal{H} \left(x-x_j\right) 
\end{flalign*}
with
\[
\begin{aligned}
&\tau = -\left(\sqrt{2} GA \Xi_1 \Theta_1 \Theta_2 \right)^{-1}\\
&\Upsilon_1 = \left(-a \left(EI \rho A \omega ^2+2 (GA)^2- GA \rho I \omega ^2\right)+\Xi _1\right)\\
&\Upsilon_2 = \left(a \left(EI \rho A \omega ^2+2 (GA)^2- GA \rho I \omega ^2\right)+\Xi _1\right)\\
&\Theta_1 = \left[\Xi_1(a EI~ GA)^{-1}-p_1\right]^{1/2}\\
&\Theta_2 = \left[-\left(\Xi_1(a EI~ GA)^{-1}+p_1\right)\right]^{1/2}\\
&\Xi_1 = \left(a^2 \omega ^2 \left((EI~\rho A \omega)^2+2 EI~GA~\rho A \left(2 GA- \rho I \omega ^2\right) \right. \right.\\ 
& \left. \left. + (GA \rho I \omega)^2\right) -2 a EI k_{eq} \left(EI \rho A \omega ^2 \right.\right.\\
&\left.\left. +2 (GA)^2-GA \rho I \omega ^2\right)+(EI k_{eq})^2\right)^{1/2}
\end{aligned}
\]
Finally, \eqref{eqn:pi} become:
\begin{flalign}\label{eqn:pi_distr}
&\Pi_{k} = \omega_{k}^{-2}\mu(\omega_k)\int_{0}^{L}a^{-1}V_{k}^{2}(x_j)~\text{d}x + 2\rho A \int_{0}^{L}V_{k}^{2}(x)~\text{d}x\nonumber\\ 
&+ 2\rho I \int_{0}^{L}\Phi_{k}^{2}(x)~\text{d}x &&
\end{flalign}

Eq.~\eqref{eqn:p1_p2_distr} through Eq.~\eqref{eqn:pi_distr} can readily be obtained considering that, when the resonators are modelled as exerting distributed forces over the mutual distance $a$ \cite{chen2011containing,chen2011dynamic}, the equations of motion \eqref{eqn:disp_freq}-\eqref{eqn:rot_freq} in the frequency domain revert to:

\begin{flalign}
&GA\left(\gmtder{2}{V}{x}+\gtder{\Phi}{x}\right)+\rho A \omega^2 V - \frac{k_{eq}(\omega)}{a}V + f_{v} = 0 \label{eqn:disp_freq_distr} &&\\
&EI\gmtder{2}{\Phi}{x}- GA\left(\gtder{V}{x}+\Phi\right)+\rho I \omega^2 \Phi + f_{\phi}= 0 && \label{eqn:rot_freq_distr}
\end{flalign}
where $a$ is the mutual distance of the resonators and $k_{eq}(\omega)$ is the frequency-dependent stiffness \eqref{eqn:equivalent_DS} of the resonator. Corresponding changes to the equations of motion \eqref{eqn:const_1}-\eqref{eqn:const_2} in the time domain are straightforward and not reported for brevity.

\bibliographystyle{unsrt}

\bibliography{sandwich_beams}

\begin{thebibliography}{10}

\bibitem{xiao2012broadband}
Yong Xiao, Jihong Wen, and Xisen Wen.
\newblock Broadband locally resonant beams containing multiple periodic arrays
  of attached resonators.
\newblock {\em Physics Letters A}, 376(16):1384--1390, 2012.

\bibitem{xiao2013flexural}
Yong Xiao, Jihong Wen, Dianlong Yu, and Xisen Wen.
\newblock Flexural wave propagation in beams with periodically attached
  vibration absorbers: band-gap behavior and band formation mechanisms.
\newblock {\em Journal of Sound and Vibration}, 332(4):867--893, 2013.

\bibitem{xiao2013theoretical}
Yong Xiao, Jihong Wen, Gang Wang, and Xisen Wen.
\newblock Theoretical and experimental study of locally resonant and bragg band
  gaps in flexural beams carrying periodic arrays of beam-like resonators.
\newblock {\em Journal of Vibration and Acoustics}, 135(4), 2013.

\bibitem{sun2010theory}
Hongwei Sun, Xingwen Du, and P~Frank Pai.
\newblock Theory of metamaterial beams for broadband vibration absorption.
\newblock {\em Journal of Intelligent Material Systems and Structures},
  21(11):1085--1101, 2010.

\bibitem{zhu2014chiral}
R~Zhu, XN~Liu, GK~Hu, CT~Sun, and GL~Huang.
\newblock A chiral elastic metamaterial beam for broadband vibration
  suppression.
\newblock {\em Journal of Sound and Vibration}, 333(10):2759--2773, 2014.

\bibitem{pai2010metamaterial}
P~Frank Pai.
\newblock Metamaterial-based broadband elastic wave absorber.
\newblock {\em Journal of Intelligent Material Systems and Structures},
  21(5):517--528, 2010.

\bibitem{hu2018internally}
Guobiao Hu, Lihua Tang, and Raj Das.
\newblock Internally coupled metamaterial beam for simultaneous vibration
  suppression and low frequency energy harvesting.
\newblock {\em Journal of Applied Physics}, 123(5):055107, 2018.

\bibitem{casalotti2018metamaterial}
Arnaldo Casalotti, Sami El-Borgi, and Walter Lacarbonara.
\newblock Metamaterial beam with embedded nonlinear vibration absorbers.
\newblock {\em International Journal of Non-Linear Mechanics}, 98:32--42, 2018.

\bibitem{wang2016multi}
Ting Wang, Mei-Ping Sheng, and Qing-Hua Qin.
\newblock Multi-flexural band gaps in an euler--bernoulli beam with lateral
  local resonators.
\newblock {\em Physics Letters A}, 380(4):525--529, 2016.

\bibitem{chen2011containing}
Jung-San Chen, B~Sharma, and CT~Sun.
\newblock Dynamic behaviour of sandwich structure containing spring-mass
  resonators.
\newblock {\em Composite Structures}, 93(8):2120--2125, 2011.

\bibitem{hussein2014dynamics}
Mahmoud~I Hussein, Michael~J Leamy, and Massimo Ruzzene.
\newblock Dynamics of phononic materials and structures: Historical origins,
  recent progress, and future outlook.
\newblock {\em Applied Mechanics Reviews}, 66(4), 2014.

\bibitem{sugino2017general}
Christopher Sugino, Yiwei Xia, Stephen Leadenham, Massimo Ruzzene, and Alper
  Erturk.
\newblock A general theory for bandgap estimation in locally resonant
  metastructures.
\newblock {\em Journal of Sound and Vibration}, 406:104--123, 2017.

\bibitem{baravelli2013internally}
Emanuele Baravelli and Massimo Ruzzene.
\newblock Internally resonating lattices for bandgap generation and
  low-frequency vibration control.
\newblock {\em Journal of Sound and Vibration}, 332(25):6562--6579, 2013.

\bibitem{beli2018wave}
D~Beli, JRF Arruda, and M~Ruzzene.
\newblock Wave propagation in elastic metamaterial beams and plates with
  interconnected resonators.
\newblock {\em International Journal of Solids and Structures}, 139:105--120,
  2018.

\bibitem{zhou2019actively}
Weijian Zhou, Weiqiu Chen, Zhenyu Chen, CW~Lim, et~al.
\newblock Actively controllable flexural wave band gaps in beam-type acoustic
  metamaterials with shunted piezoelectric patches.
\newblock {\em European Journal of Mechanics-A/Solids}, 77:103807, 2019.

\bibitem{liu2020study}
Panxue Liu, Shuguang Zuo, Xudong Wu, Lingzhou Sun, and Qi~Zhang.
\newblock Study on the vibration attenuation property of one finite and hybrid
  piezoelectric phononic crystal beam.
\newblock {\em European Journal of Mechanics-A/Solids}, page 104017, 2020.

\bibitem{krushynska2017coupling}
AO~Krushynska, Marco Miniaci, Federico Bosia, and NM~Pugno.
\newblock Coupling local resonance with bragg band gaps in single-phase
  mechanical metamaterials.
\newblock {\em Extreme Mechanics Letters}, 12:30--36, 2017.

\bibitem{miniaci2016spider}
Marco Miniaci, Anastasiia Krushynska, Alexander~B Movchan, Federico Bosia, and
  Nicola~M Pugno.
\newblock Spider web-inspired acoustic metamaterials.
\newblock {\em Applied Physics Letters}, 109(7):071905, 2016.

\bibitem{chen2011dynamic}
Jung-San Chen and CT~Sun.
\newblock Dynamic behavior of a sandwich beam with internal resonators.
\newblock {\em Journal of Sandwich Structures \& Materials}, 13(4):391--408,
  2011.

\bibitem{sharma2016local}
Bhisham Sharma and Chin-Teh Sun.
\newblock Local resonance and bragg bandgaps in sandwich beams containing
  periodically inserted resonators.
\newblock {\em Journal of Sound and Vibration}, 364:133--146, 2016.

\bibitem{sharma2016impact}
B~Sharma and CT~Sun.
\newblock Impact load mitigation in sandwich beams using local resonators.
\newblock {\em Journal of Sandwich Structures \& Materials}, 18(1):50--64,
  2016.

\bibitem{chen2012reducing}
Jung-San Chen and CT~Sun.
\newblock Reducing vibration of sandwich structures using antiresonance
  frequencies.
\newblock {\em Composite Structures}, 94(9):2819--2826, 2012.

\bibitem{chen2016sandwich}
Jung-San Chen and Song-Mao Tsai.
\newblock Sandwich structures with periodic assemblies on elastic foundation
  under moving loads.
\newblock {\em Journal of Vibration and Control}, 22(10):2519--2529, 2016.

\bibitem{chen2013wave}
Jung-San Chen and CT~Sun.
\newblock Wave propagation in sandwich structures with resonators and periodic
  cores.
\newblock {\em Journal of Sandwich Structures \& Materials}, 15(3):359--374,
  2013.

\bibitem{guo2017flexural}
Zhiwei Guo, Meiping Sheng, and Jie Pan.
\newblock Flexural wave attenuation in a sandwich beam with viscoelastic
  periodic cores.
\newblock {\em Journal of Sound and Vibration}, 400:227--247, 2017.

\bibitem{li2020phononic}
Jingru Li, Peng Yang, and Sheng Li.
\newblock Phononic band gaps by inertial amplification mechanisms in periodic
  composite sandwich beam with lattice truss cores.
\newblock {\em Composite Structures}, 231:111458, 2020.

\bibitem{li2019novel}
Bing Li, Yongquan Liu, and Kwek-Tze Tan.
\newblock A novel meta-lattice sandwich structure for dynamic load mitigation.
\newblock {\em Journal of Sandwich Structures \& Materials}, 21(6):1880--1905,
  2019.

\bibitem{williams1970automatic}
FW~Williams and WH~Wittrick.
\newblock An automatic computational procedure for calculating natural
  frequencies of skeletal structures.
\newblock {\em International Journal of Mechanical Sciences}, 12(9):781--791,
  1970.

\bibitem{qi2004accurate}
Zhaohui Qi, David Kennedy, and Frederic~Ward Williams.
\newblock An accurate method for transcendental eigenproblems with a new
  criterion for eigenfrequencies.
\newblock {\em International Journal of Solids and Structures},
  41(11-12):3225--3242, 2004.

\bibitem{sakurai2003projection}
Tetsuya Sakurai and Hiroshi Sugiura.
\newblock A projection method for generalized eigenvalue problems using
  numerical integration.
\newblock {\em Journal of computational and applied mathematics},
  159(1):119--128, 2003.

\bibitem{asakura2009numerical}
Junko Asakura, Tetsuya Sakurai, Hiroto Tadano, Tsutomu Ikegami, and Kinji
  Kimura.
\newblock A numerical method for nonlinear eigenvalue problems using contour
  integrals.
\newblock {\em JSIAM Letters}, 1:52--55, 2009.

\bibitem{ikegami2010filter}
Tsutomu Ikegami, Tetsuya Sakurai, and Umpei Nagashima.
\newblock A filter diagonalization for generalized eigenvalue problems based on
  the sakurai--sugiura projection method.
\newblock {\em Journal of Computational and Applied Mathematics},
  233(8):1927--1936, 2010.

\bibitem{falsone2002use}
G~Falsone.
\newblock The use of generalised functions in the discontinuous beam bending
  differential equations.
\newblock {\em International Journal of Engineering Education}, 18(3):337--343,
  2002.

\bibitem{caddemi2013exact}
S~Caddemi and I~Cali{\`o}.
\newblock The exact explicit dynamic stiffness matrix of multi-cracked
  euler--bernoulli beam and applications to damaged frame structures.
\newblock {\em Journal of Sound and Vibration}, 332(12):3049--3063, 2013.

\bibitem{biondi2007euler}
B~Biondi and S~Caddemi.
\newblock Euler--bernoulli beams with multiple singularities in the flexural
  stiffness.
\newblock {\em European Journal of Mechanics-A/Solids}, 26(5):789--809, 2007.

\bibitem{burlon2016exact}
Andrea Burlon, Giuseppe Failla, and Felice Arena.
\newblock Exact frequency response analysis of axially loaded beams with
  viscoelastic dampers.
\newblock {\em International Journal of Mechanical Sciences}, 115:370--384,
  2016.

\bibitem{di2018flexural}
Salvatore Di~Lorenzo, Christoph Adam, Andrea Burlon, Giuseppe Failla, and
  Antonina Pirrotta.
\newblock Flexural vibrations of discontinuous layered elastically bonded
  beams.
\newblock {\em Composites Part B: Engineering}, 135:175--188, 2018.

\bibitem{wang2007vibration}
Jialai Wang and Pizhong Qiao.
\newblock Vibration of beams with arbitrary discontinuities and boundary
  conditions.
\newblock {\em Journal of Sound and Vibration}, 308(1-2):12--27, 2007.

\bibitem{failla2016exact}
Giuseppe Failla.
\newblock An exact generalised function approach to frequency response analysis
  of beams and plane frames with the inclusion of viscoelastic damping.
\newblock {\em Journal of Sound and Vibration}, 360:171--202, 2016.

\bibitem{liu2007design}
Yaozong Liu, Dianlong Yu, Li~Li, Honggang Zhao, Jihong Wen, and Xisen Wen.
\newblock Design guidelines for flexural wave attenuation of slender beams with
  local resonators.
\newblock {\em Physics Letters A}, 362(5-6):344--347, 2007.

\bibitem{bestle2014recursive}
Dieter Bestle, Laith Abbas, and Xiaoting Rui.
\newblock Recursive eigenvalue search algorithm for transfer matrix method of
  linear flexible multibody systems.
\newblock {\em Multibody System Dynamics}, 32(4):429--444, 2014.

\bibitem{kawashima1984vibration}
Sukeo Kawashima and T~Fujimoto.
\newblock Vibration analysis of frames with semi-rigid connections.
\newblock {\em Computers \& Structures}, 19(1-2):85--92, 1984.

\bibitem{sakurai2013efficient}
Tetsuya Sakurai, Yasunori Futamura, and Hiroto Tadano.
\newblock Efficient parameter estimation and implementation of a contour
  integral-based eigensolver.
\newblock {\em Journal of Algorithms \& Computational Technology},
  7(3):249--269, 2013.

\bibitem{failla2020exact}
Giuseppe Failla, Roberta Santoro, Andrea Burlon, and Andrea~Francesco Russillo.
\newblock An exact approach to the dynamics of locally-resonant beams.
\newblock {\em Mechanics Research Communications}, 103:103460, 2020.

\bibitem{adam2017moving}
Christoph Adam, Salvatore Di~Lorenzo, Giuseppe Failla, and Antonina Pirrotta.
\newblock On the moving load problem in beam structures equipped with tuned
  mass dampers.
\newblock {\em Meccanica}, 52(13):3101--3115, 2017.

\bibitem{oliveto1997complex}
G~Oliveto, Adolfo Santini, and E~Tripodi.
\newblock Complex modal analysis of a flexural vibrating beam with viscous end
  conditions.
\newblock {\em Journal of Sound and Vibration}, 200(3):327--345, 1997.

\bibitem{MATLAB:2010}
MATLAB.
\newblock {\em version 7.10.0 (R2010a)}.
\newblock The MathWorks Inc., Natick, Massachusetts, 2010.

\end{thebibliography}

\end{document}